\documentclass[aps,prl,twocolumn,superscriptaddress,showpacs,10pt,longbibliography]{revtex4-2}
\usepackage{amsmath}
\usepackage[dvips,xdvi]{graphicx}
\usepackage{amssymb}
\usepackage{epsfig}
\usepackage{xcolor}
\usepackage{physics}
\usepackage{soul}
\usepackage[utf8]{inputenc}
\usepackage[colorlinks=true,citecolor=blue,urlcolor=magenta,linkcolor=blue]{hyperref}

\begin{document}
	\title{Lattice-enabled detection of spin-dependent three-body interactions} 
	\author{C. Binegar}
	\author{J. O. Austin-Harris}
    \affiliation{Department of Physics, Oklahoma State University, Stillwater, Oklahoma 74078, USA}
	\author{S. E. Begg}
	\affiliation{Department of Physics, Oklahoma State University, Stillwater, Oklahoma 74078, USA}
    \affiliation{Department of Physics, The University of Texas at Dallas, Richardson, Texas 75080, USA}
	\author{P. Sigdel}       
	\author{T. Bilitewski}
	\email{thomas.bilitewski@okstate.edu}
	\author{Y. Liu}
	\email{yingmei.liu@okstate.edu}
	\affiliation{Department of Physics, Oklahoma State University, Stillwater, Oklahoma 74078, USA}
	\date{\today}
	
    \begin{abstract}
    	We present the experimental detection of coherent three-body interactions, often masked by stronger two-body effects, through nonequilibrium spin dynamics induced by controllably quenching lattice-confined spinor gases. Three-body interactions are characterized through both real-time and frequency domain analyses of the observed dynamics.  Our results, well-described by an extended Bose-Hubbard model, further demonstrate the importance of three-body interactions for correctly determining atom distributions in lattice systems, which has applications in quantum sensing via spin singlets. The techniques demonstrated in this work can be directly applied to other atomic species, offering a promising avenue for future studies of higher-body interactions with broad relevance to strongly-interacting quantum systems.
	\end{abstract}
    
	\maketitle

Higher-body interactions involving three or more particles arise within effective descriptions of the low-energy physics of confined quantum systems \cite{Mahmud2013Dynamics,Johnson2009,Tiesinga2011,Johnson_2012,PhysRevA.90.043631,Perlin_2019,PhysRevA.90.041602}, while fundamental interactions are typically of two-body type. A range of platforms, including neutral atom Bose-Einstein condensates (BECs)~\cite{Will2010,ma2011photon,mark2011precision}, atomic clocks~\cite{Goban2018Emergence}, and Rydberg systems~\cite{Gambetta2020Engineering,Gambetta2020LongRange,Fey2019Effective}, have experimentally demonstrated these effective higher-body interactions. 
This offers a promising route towards quantum simulation of high-energy physics models and lattice gauge theories \cite{Schweizer_2019,Banuls_2020,PhysRevX.3.041018,PhysRevLett.118.070501}, exotic spin \cite{Motrunich2005Variational,Andrade2022,PhysRevLett.93.056402} and (topological) many-body phases \cite{Buchler_2007,PhysRevLett.101.150405,Harshman_2020,PhysRevA.94.063610}, and three-body phenomena such as the nuclear Fujita-Miyazawa force~\cite{Honda2025Exploring,Endo2025ThreeBody} and Efimov physics~\cite{Braaten2007Efimov,Honda2025Exploring,Secker2021Multichannel,Musolino2022,Endo2025ThreeBody,Kraats2024Emergent}. These higher-body interactions also facilitate the fast generation of highly-entangled states \cite{Cieslinski_2023,PhysRevLett.107.260502}, and may enhance quantum computation approaches across multiple platforms~\cite{Dai2017FourBody,Menke2022Demonstration,Katz2023Demonstration,Nilss2025Resonant} where they enable multi-qubit gates \cite{Wang2001,Ezawa2024Systematic} and error correction protocols \cite{RevModPhys.87.307,Kitaev2003,Vy2013ErrorTransparent}. 

In the context of systems with a spin degree of freedom, interactions generically depend on the internal state, and therefore can be density or spin dependent. One such platform is lattice-confined spinor gases, which provides a programmable large-scale many-body simulator with control over parameters including dimensionality and interactions~\cite{Kawaguchi_2012,Stamper2013,Zihe2019,Jared2021Manipulation,Jared2021Quantum,Zach1,Jared2024,Jie2016,Zach2,Mahmud2013Dynamics,Nabi2022Interplay,nabi2018quantum}, and is particularly well suited to study spin-dependent higher-body interactions~\cite{Mahmud2013Dynamics,Nabi2022Interplay,nabi2018quantum}. While these interactions are predicted to be important for many phenomena, they have remained experimentally underexplored due to technical challenges~\cite{Mahmud2013Dynamics,Nabi2022Interplay,nabi2018quantum,hincapie2018spin,Mestrom2021Mixing}.

\begin{figure}[t!]
    \includegraphics[width=86mm]{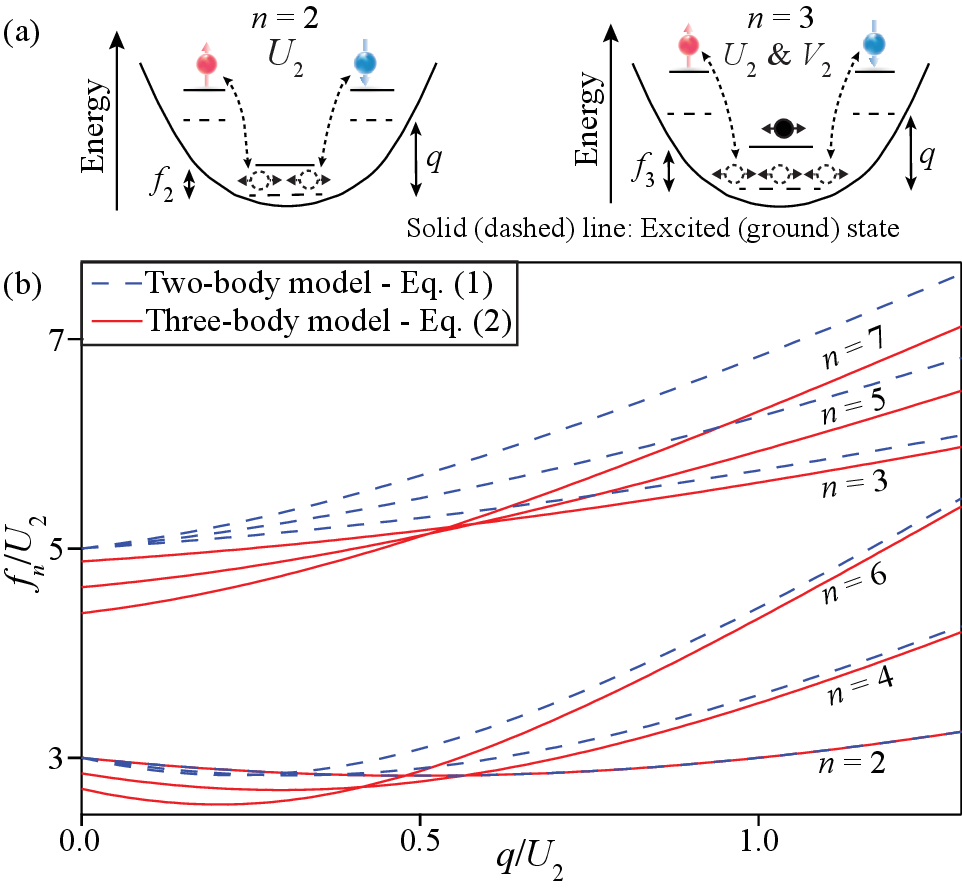}
    \caption{
    (a) Illustration of $F=1$ atoms in lattice sites with a filling factor $n=2$ (left) or $n=3$ (right) based on the three-body model, Eq.~\eqref{ThreeBody}. Black filled circles (open circles) represent $m_F=0$ atoms in the excited (ground) states. Red (blue) spheres represent $m_F=1$ ($m_F=-1$) atoms in excited states. The characteristic frequency of sites with $n$ particles is $f_n = \Delta E_n/h$ where $\Delta E_n$ is the energy gap between the first excited state and the ground state and $h$ is the Planck constant. $V_2$ ($U_2$) is the spin-dependent three-body (two-body) interaction. Axes are not to scale. (b) Dashed and solid lines respectively display the predicted $f_n$ versus $q$ at $V_2/U_2= -0.074$ based on the two-body and three-body models (see Eqs.~\eqref{TwoBody} and \eqref{ThreeBody}). }
    \label{fig:1}
\end{figure} 

In this Letter, we demonstrate a technique for precisely detecting spin-dependent many-body interactions via nonequilibrium spin dynamics induced by a quantum quench of lattice-confined spinor gases. Signatures of higher-body interactions, typically masked by the much stronger two-body effects, are revealed through carefully-designed experimental sequences that controllably quench the many-body interactions or quadratic Zeeman energy in deep lattices. The observed influence of three-body interactions agrees with predictions for a spherical harmonic lattice-trapping potential using an extended Bose-Hubbard model. The detected nonequilibrium spin dynamics manifest as multiple Rabi-type oscillations in the spin populations that, when studied via frequency analysis, can also be used to probe number and spatial distributions of three-dimensional (3D) lattice systems~\cite{Zihe2019,Jared2021Manipulation}. Our results demonstrate the importance of three-body interactions in dense lattice systems, especially for correctly resolving atom distributions in strongly-interacting quantum systems. This has applications to quantum sensing~\cite{Sun2017Efficient,Toth2010Generation,urizar2013macroscopic,behbood2014generation,Evrard2021Observation} and lays the foundation for studies of higher-body spin-dependent interactions. 

{\it Experimental Sequence---} 
Each experimental cycle begins by generating an $F=1$ spinor BEC of up to $1\times 10^5$ sodium atoms in their superfluid ground state, the longitudinal polar (LP) state where $\rho_0 = 1$ and $M = 0$. Here $\rho_{m_F}$ is the fractional population of the $m_F$ hyperfine state and $M = \rho_1 - \rho_{-1}$ is the magnetization. The atoms are then loaded into a cubic optical lattice. Nonequilibrium spin dynamics are initiated by one of two quantum quench sequences which are characterized by either a sudden change of lattice depth or the magnetic field (see Supplemental Materials (SM)~\cite{SM}). These sequences result in a quench of the inter-particle interactions or the quadratic Zeeman shift $q$, respectively.
The atoms are then allowed to evolve in the lattice for a time $t_\mathrm{hold}$ before the spin populations are measured with a microwave imaging method~\cite{Zihe2019,Jared2021Manipulation,Jared2021Quantum}.

{\it Model---} 
Deep lattices, by tightly confining atoms into individual lattice sites as well as precisely tuning the interactions, enable the study of few-body nonequilibrium dynamics. Spin dynamics of spinor gases are driven by the competition between the quadratic Zeeman shift $q$ and spin-dependent interactions, for example, the two-body spin-dependent interaction $U_2$~\cite{Kawaguchi_2012,Stamper2013,Lichao2014,Lichao2015,Jie2014,Zihe2019,Jared2021Quantum,Jared2021Manipulation,Zach1,Jared2024}. A typical example for a doubly occupied lattice site $(n=2)$ is shown on the left side of Fig.~\ref{fig:1}(a): a four-wave interconversion of a pair of $m_F=0$ atoms to/from one $m_F=1$ atom and one $m_F=-1$ atom mediated by $U_2$ and $q$. The sign of $U_2$ is determined by the atomic species, for sodium ($^{23}$Na) as in our experimental system $U_2>0$~\cite{Stamper2013,Mahmud2013Dynamics,Lichao2014,Zihe2019,Jared2021Manipulation}.
 In principle, any system with more than two particles also has higher-body interactions, for example, the three-body spin-dependent interaction $V_2$, although models that include only two-body interactions are often sufficient to describe dilute systems, such as atoms in free space or in a shallow lattice~\cite{Hammond2022Tunable,Jie2016,Zach1,Zach2,Jared2024}. In deep lattices, which greatly enhance the atomic density, higher-body interactions may become more pronounced and detectable. A diagram for a triply occupied lattice site $(n=3)$ is shown on the right side of Fig.~\ref{fig:1}(a): a similar four-wave interconversion involving 3-particles occurs, and $\Delta E_n$, the energy gaps between the first excited state and ground state for sites with $n$ particles, are modified by $V_2$~\cite{Mahmud2013Dynamics}.  Additional but smaller corrections in the form of effective $n$-body interactions occur for sites with occupations $n>3$~\cite{Mahmud2013Dynamics}.
 
\begin{figure}[t]
    \includegraphics[width=86mm]{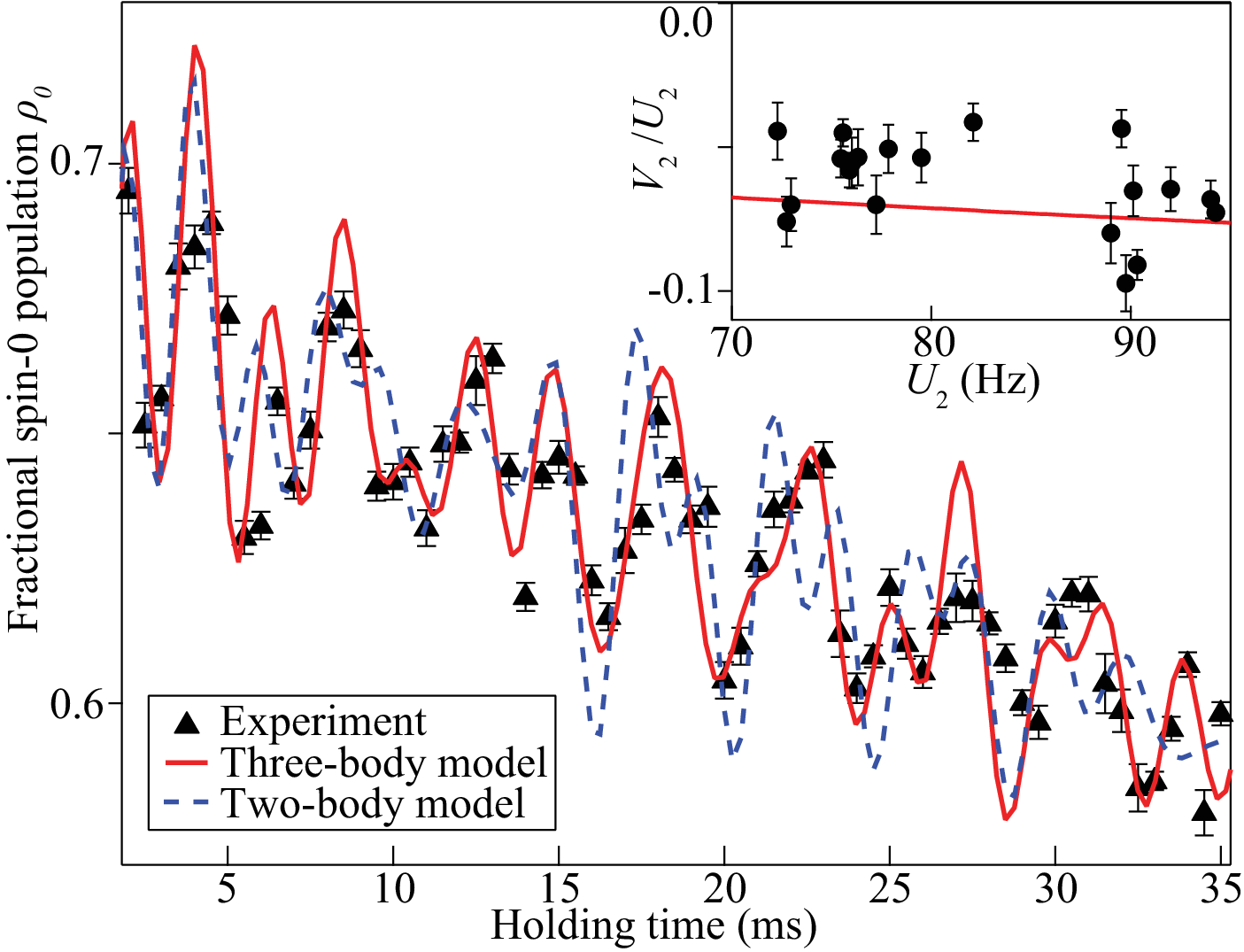}
    \caption{\textit{Real-time Analysis.} Triangles display the observed time evolution of $\rho_0$ at $q=75~\mathrm{Hz}$ and $U_2\approx73~\mathrm{Hz}$. The solid (dashed) line is a multi-sinusoidal fit with frequencies $f_n$ predicted by the three-body (two-body) model (see Eqs.~\eqref{TwoBody}-\eqref{ThreeBody}). Inset: Circles represent the ratio of $V_2$ to $U_2$ extracted from fitting the observed spin dynamics (see the solid line in main panel and Eq.~\eqref{ThreeBody}). The line is the Eq.~\eqref{V2U2} prediction. }
    \label{fig:2}
\end{figure} 

Spin dynamics of atoms confined in deep lattices, where tunneling is negligible, can be understood via a site-independent single-site Bose-Hubbard Hamiltonian $\hat{H}_3$ ($\hat{H}_2$) which respectively includes (neglects) three-body interactions~\cite{Mahmud2013Dynamics,nabi2018quantum,Zihe2019,hincapie2018spin,Nabi2022Interplay}:
\begin{align}
		\hat{H}_2/h=& {\frac{U_0}{2} \hat{n}(\hat{n}-1) + \frac{U_2}{2}(\vec{S}^2-2\hat{n}}) + q (\hat{n}_1+\hat{n}_{-1})\label{TwoBody},\\
        \hat{H}_3/h =& \hat{H}_{2}/h+\frac{V_0}{6}\hat{n}(\hat{n}-1)(\hat{n}-2) + \frac{V_2}{6}(\vec{S}^2-2\hat{n}) (\hat{n}-2)\label{ThreeBody}.
\end{align}
Here $V_0$ ($U_0$) is the spin-independent three-body (two-body) interaction, $\hat{n}=\sum_{m_F} \hat{n}_{m_F}$ is the number operator, $\vec{S}$ is the spin operator, and $h$ is the Planck constant. Figure~\ref{fig:1}(b) shows a typical comparison of the characteristic energy gaps for sites of $n\leq7$ as determined by $\hat{H}_2$ and $\hat{H}_3$ for a sodium system. Note that the atom number $n$ is conserved in the isolated site limit of a deep lattice, and the pure density-interactions ($U_0$, $V_0$) are thus irrelevant to the dynamics studied in this Letter.   

{\it Spin Oscillations---} 
The observed post-quench spin dynamics (see Fig.~\ref{fig:2}) can be described by a sum of multiple Rabi-type oscillations with characteristic frequencies $f_n=\Delta E_n/h$ that correspond to the energy gap $\Delta E_n$~\cite{Zihe2019,Jared2021Manipulation}. The inclusion of three-body terms in Eq.~\eqref{ThreeBody} alters the structure of eigenstates and the relevant energy gaps, resulting in distinct predictions for $f_n$ between the three-body model (solid lines) and the two-body model (dashed lines), as seen in Fig.~\ref{fig:1}(b).   
Because independently measuring the number distribution of 3D lattice-confined spinor gases is technically challenging~\cite{Jared2021Manipulation}, determining which frequency signature in the dynamics corresponds to a given $n$ is reliant on comparisons to the theoretical model. To conclusively demonstrate the presence of three-body effects in the dynamics, an observed frequency component must therefore be well separated from all characteristic two-body frequencies predicted by Eq.~\eqref{TwoBody}, in addition to being close to a characteristic three-body frequency predicted by Eq.~\eqref{ThreeBody}.  This restriction leads to two regimes where signatures of three-body effects are predicted to be experimentally observable (see Fig.~\ref{fig:1}(b)). One regime is in the limit of very small $q$, where the spin-dependent interactions dominates the dynamics and the two-body model predicts degenerate frequencies $f_n=3U_2$ ($5U_2$) for all sites with even (odd) occupation. This degeneracy is lifted by three-body effects resulting in distinct oscillation frequencies for each $n$ and giving a clear experimental signature for the presence of three-body interactions. The other regime lies at $q > 0.5U_2$, where sites of odd occupation are predicted to show strong signatures of three-body effects in their characteristic frequencies $f_n$. For intermediate $q$ (i.e.,  $q/U_2\sim0.5$), the characteristic frequencies derived from the three-body model appear to be experimentally indistinguishable from nearby frequencies derived from the two-body model (see Fig.~\ref{fig:1}(b)). 

{\it Extraction of interaction strengths---} 
The characteristic frequencies $f_n$ at a given $q$ are determined by the three-body interaction $V_2$ and two-body interaction $U_2$ (see Eq.~\eqref{ThreeBody} and Fig.~\ref{fig:1}). The strength of $V_2$ and $U_2$ can therefore be extracted through fitting the observed spin dynamics with a multi-sinusoidal function with the two interaction terms left as free parameters~\cite{Zihe2019}. 
A typical example is shown in Fig.~\ref{fig:2}: a fit using frequencies derived from the two-body model (blue dashed line) fails to capture multiple features of the spin oscillations that are well described by a similar fit that uses frequencies derived from the three-body model (red solid line), demonstrating that three-body interactions contribute to the observed spin dynamics. We repeat this time-trace analysis for twenty different sets of parameters and the extracted strengths of $V_2$ and $U_2$ over a wide range of conditions are shown in the inset of Fig.~\ref{fig:2}.
\begin{figure}[t]		
\includegraphics[width=86mm]{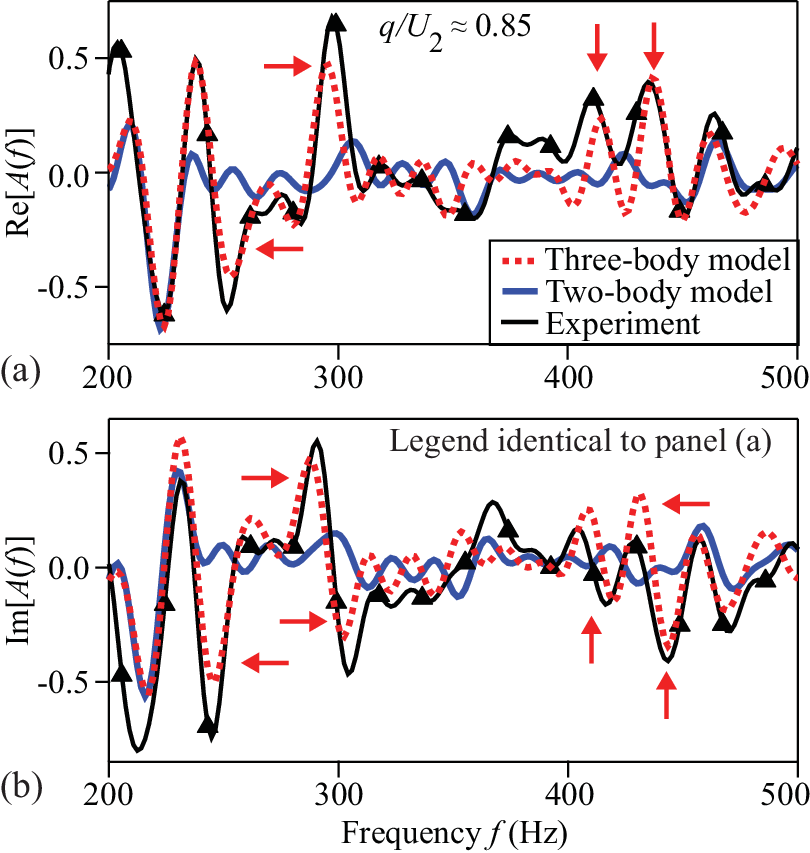}
		\caption{ \textit{Frequency-Domain Analysis.}
        Black triangles (black thin solid lines) display the (a) real and (b) imaginary components of the non-padded (zero-padded) Fourier spectrum $A(f)$ for the observed spin dynamics at $q/U_2\approx 0.85$. Red dotted (blue thick solid) lines show theoretical simulations of the Fourier spectrum based on the three-body (two-body) model at the optimal  
        $V_2=-4.8(6)~\mathrm{Hz}$ and $U_2=76(1)~\mathrm{Hz}$ (see Eq.~\eqref{TwoBody} and Eq.~\eqref{ThreeBody}).
        Red arrows highlight significantly improved agreements in the theory-experiment comparison using the three-body model.
        \label{fig:3}
        }
\end{figure}

While the three-body density interaction $V_0$ does not affect the observed spin dynamics, it may, in principle, be obtained from the extracted $V_2$ because $V_2/V_0$ is predicted to be equal to $2 U_2/U_0$~\cite{Mahmud2013Dynamics}.   For sodium, as in our system, $U_2/U_0\approx0.035$~\cite{Zihe2019}. Based on the extracted $V_2$ (see Fig.~\ref{fig:2} inset), $V_0$ is proportional to the lattice depth and within $-50~\mathrm{Hz}$ and $-150~\mathrm{Hz}$ for the data presented in this Letter. These results complement the extraction of $V_0$ accomplished previously by measuring revivals of visibility in scalar systems~\cite{Johnson2009,Will2010,Mahmud2013Dynamics}. The inset of Fig.~\ref{fig:2} shows a qualitative agreement between the experimental $V_2/U_2$ and the prediction (solid line) for an isotropic spherical harmonic lattice trapping potential as expressed below: 
    \begin{equation}
        \frac{V_2}{U_2}=-1.34  \frac{hU_0}{\sqrt{u_L E_R}}.\label{V2U2}
    \end{equation}
Here $u_L$ is the lattice depth, $E_R$ is the recoil energy, and $V_0=-0.67\frac{hU_0^2}{\sqrt{u_L E_R}}$~\cite{Johnson2009,Tiesinga2011,Mahmud2013Dynamics}. The quantitative theory-experiment disagreement could be a result of the anisotropy present in our lattice system, which has been predicted to result in deviations of up to $10\%$ in experiments~\cite{Johnson2009}. 

{\it Spectra---}  
In addition to the time-based analysis, we can also study the observed spin oscillations in frequency space.  Figure~\ref{fig:3}(a) and \ref{fig:3}(b) respectively display the real and imaginary components of the Fourier spectrum $A(f) = \mathcal{F}[\rho_0(t)]$, where $\mathcal{F}$ is the Fourier transform operator, of the observed dynamics (black lines and markers) at $q/U_2\approx0.85$ compared to simulations of the full exact time dynamics performed using the two-body (blue lines) and three-body (red lines) Hamiltonians. The shown theory results are obtained for `optimal' fit values of $U_2$ and $V_2$ determined by minimizing the distance $D$ between the theory and experiment spectra (See SM for details~\cite{SM}). The $V_2$ and $U_2$ values obtained from this method are found to be consistent within errors with the values obtained from the time trace analysis shown in the inset of Fig.~\ref{fig:2}. Since the density measurement contains contributions from all lattice sites of $n>1$~\cite{Zihe2019,Jared2021Manipulation,SM}, this procedure also involves optimizing over $\chi_n$, the unknown fraction of atoms occupying sites of a given $n$. Our results show that the observed spin dynamics often cannot be explained by simulations solely based on the two-body model, which misses prominent spectral peaks that are captured by the three-body model (see areas highlighted by red arrows in Fig.~\ref{fig:3}). Furthermore, the agreement visible in both the  real and imaginary components indicates that the three-body model not only approximates the peak locations, but also obtains the correct phase. The phase agreement is nontrivial as it involves the eigenvectors and their overlap with the initial state \cite{SM}, suggesting that the \textit{ab}-\textit{initio} modeling captures key aspects of the system dynamics. In some instances, the two-body model can give the wrong phase even after accounting for the shift in peak frequency \cite{SM}.
These features provide additional experimental signatures  for the presence of the three-body interactions, supporting the time-trace analysis in Fig.~\ref{fig:2}.

\begin{figure}[t]
		\includegraphics[width=86mm]{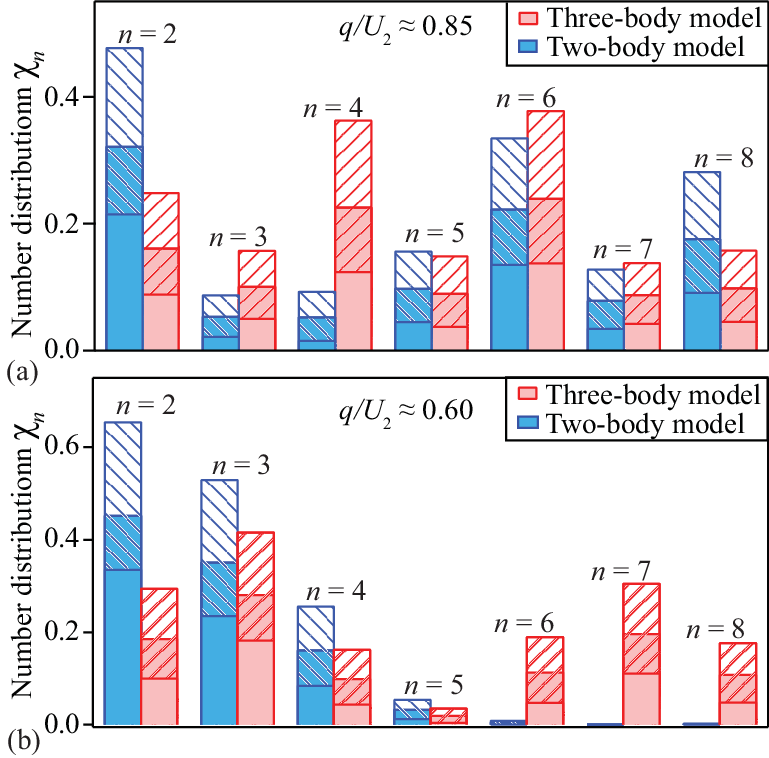}
		\caption{
      Number distribution $\chi_n$ for results extracted using the two-body (blue) and three-body models (red) for the (a) $q/U_2 \approx 0.85$ dataset shown in Fig.~\ref{fig:3} and (b) a dataset taken at $q/U_2 \approx 0.60$. Hatched regions indicate the uncertainties (see SM~\cite{SM}). 
        \label{fig:4}
        }	
	\end{figure}

{\it Number distributions---}   
The inclusion or exclusion of three-body interactions can have drastic effects on the number distribution $\chi_n$ obtained from the observed spectra, as shown by results extracted at two typical conditions in Fig.~\ref{fig:4}. Because $\chi_n$ is normalized such that $\sum_{n>1} \chi_n =1$, spectral features associated with sites of a given filling factor $n$ that are missed by simulations using the two-body Hamiltonian result in larger extracted $\chi_n$ for other $n$. This effect is particularly apparent for $\chi_2$, the fraction of atoms in doubly occupied lattice sites $(n=2)$ whose associated features in the frequency spectrum are captured equally well by both models. The $\chi_2$ extracted from simulations based on the two-body model are typically significantly larger than $\chi_2$ extracted from simulations based on the three-body model (see Fig.~\ref{fig:4}) due to features associated with other populations being missed by simulations using the two-body model~\cite{SM}. Our results in Fig.~\ref{fig:4} highlight how the failure of the two-body model to describe the observed spin dynamics (see Fig.~\ref{fig:2} and Fig.~\ref{fig:3}) can distort conclusions derived from the observed spin dynamics. Accurately determining the number distribution is a crucial foundation for developing experimental techniques that can precisely detect and control singlet states, which are predicted to only form in Mott lobes with even occupation numbers~\cite{Imambekov2003Spin,Jie2016}. The creation of singlet states with long lifespans and resistance to environmental noise has numerous applications in quantum memories and quantum metrology~\cite{Toth2010Generation,Sun2017Efficient,Jie2016,Jared2021Manipulation,urizar2013macroscopic,behbood2014generation,Evrard2021Observation}.

In some previous studies~\cite{Zihe2019,Jared2021Manipulation}, the number distribution $\chi_n$ was extracted exclusively by an approach that evaluates the magnitude of the observed spectral peaks and compares their center frequency to the predicted values of $f_n$. This contrasts with our procedure discussed above, where we optimize $\chi_n$ by comparing the complex Fourier spectrum obtained from full \textit{ab}-\textit{initio} modeling of the quantum dynamics.   
Our approach utilized in this Letter avoids needing to assign peaks to a given $n$, which is nontrivial in the presence of three-body interactions where the ordering of $f_n$ can change as parameters are varied (see Fig.~\ref{fig:1}(b)). The inclusion of phase information and all many-body frequencies may also be helpful when studying sets taken closer to $q/U_2 \sim 0.5$ where the frequencies $f_n$ become indistinguishable in all but $n$ parity. For regimes where the spectral features are well-resolved, on the other hand, the approaches return distributions that are consistent within error.

{\it Conclusion---} Our work experimentally reveals and characterizes effective coherent three-body interactions in lattice-confined spinor gases. The three-body spin-dependent interaction strengths are precisely extracted via real-time and frequency domain analyses of the observed quench-induced non-equilibrium spin dynamics. We further demonstrate that inclusion of the three-body interactions is critical to accurately determine the atom number distribution in lattices. While we focus on spinor BECs of sodium atoms here, these techniques can be applied directly to other atomic species. The quantum quench method described here, in principle, also enables the detection of higher-body (beyond three-body) interactions, however, due to their smaller magnitude, technical improvements and a careful study to identify optimal parameter regimes would be required to make their detection feasible. 

The measurement of three-body interactions and determination of the atom number distribution in three-dimensional lattices build towards the wider challenge of manipulating strongly-interacting samples of cold atoms, with potential applications in interaction-based quantum enhanced sensing \cite{PhysRevA.90.041602,Pan2019HighOrder} and the generation of spin-singlet states \cite{Sun2017Efficient,Jared2021Manipulation,Evrard2021Observation,Nabi2022Interplay,nabi2018quantum}. These methods also enable the characterization of complex many-body lattice dynamics via atom number distributions, 
even in the presence of tunneling dynamics beyond the isolated site limit studied here. 
    
	\begin{acknowledgments}
		\noindent{\it Acknowledgments --} We acknowledge the support of the Noble Foundation and the National Science Foundation (NSF) through grants PHY-2207777 and PHY-2513302. Some of the computing for this project was performed at the High Performance Computing Center at Oklahoma State University supported in part through the NSF grant OAC-1531128. S.E.B. acknowledges support from the NSF through grant OMR-2228725.
	\end{acknowledgments}
    

\begin{thebibliography}{68}%
	\makeatletter
	\providecommand \@ifxundefined [1]{%
		\@ifx{#1\undefined}
	}%
	\providecommand \@ifnum [1]{%
		\ifnum #1\expandafter \@firstoftwo
		\else \expandafter \@secondoftwo
		\fi
	}%
	\providecommand \@ifx [1]{%
		\ifx #1\expandafter \@firstoftwo
		\else \expandafter \@secondoftwo
		\fi
	}%
	\providecommand \natexlab [1]{#1}%
	\providecommand \enquote  [1]{``#1''}%
	\providecommand \bibnamefont  [1]{#1}%
	\providecommand \bibfnamefont [1]{#1}%
	\providecommand \citenamefont [1]{#1}%
	\providecommand \href@noop [0]{\@secondoftwo}%
	\providecommand \href [0]{\begingroup \@sanitize@url \@href}%
	\providecommand \@href[1]{\@@startlink{#1}\@@href}%
	\providecommand \@@href[1]{\endgroup#1\@@endlink}%
	\providecommand \@sanitize@url [0]{\catcode `\\12\catcode `\$12\catcode
		`\&12\catcode `\#12\catcode `\^12\catcode `\_12\catcode `\%12\relax}%
	\providecommand \@@startlink[1]{}%
	\providecommand \@@endlink[0]{}%
	\providecommand \url  [0]{\begingroup\@sanitize@url \@url }%
	\providecommand \@url [1]{\endgroup\@href {#1}{\urlprefix }}%
	\providecommand \urlprefix  [0]{URL }%
	\providecommand \Eprint [0]{\href }%
	\providecommand \doibase [0]{https://doi.org/}%
	\providecommand \selectlanguage [0]{\@gobble}%
	\providecommand \bibinfo  [0]{\@secondoftwo}%
	\providecommand \bibfield  [0]{\@secondoftwo}%
	\providecommand \translation [1]{[#1]}%
	\providecommand \BibitemOpen [0]{}%
	\providecommand \bibitemStop [0]{}%
	\providecommand \bibitemNoStop [0]{.\EOS\space}%
	\providecommand \EOS [0]{\spacefactor3000\relax}%
	\providecommand \BibitemShut  [1]{\csname bibitem#1\endcsname}%
	\let\auto@bib@innerbib\@empty
	\bibitem [{\citenamefont {Mahmud}\ and\ \citenamefont
		{Tiesinga}(2013)}]{Mahmud2013Dynamics}%
	\BibitemOpen
	\bibfield  {author} {\bibinfo {author} {\bibfnamefont {K.~W.}\ \bibnamefont
			{Mahmud}}\ and\ \bibinfo {author} {\bibfnamefont {E.}~\bibnamefont
			{Tiesinga}},\ }\bibfield  {title} {\bibinfo {title} {Dynamics of spin-1
			bosons in an optical lattice: Spin mixing, quantum-phase-revival
			spectroscopy, and effective three-body interactions},\ }\href
	{https://doi.org/10.1103/PhysRevA.88.023602} {\bibfield  {journal} {\bibinfo
			{journal} {Phys. Rev. A}\ }\textbf {\bibinfo {volume} {88}},\ \bibinfo
		{pages} {023602} (\bibinfo {year} {2013})}\BibitemShut {NoStop}%
	\bibitem [{\citenamefont {Johnson}\ \emph {et~al.}(2009)\citenamefont
		{Johnson}, \citenamefont {Tiesinga}, \citenamefont {Porto},\ and\
		\citenamefont {Williams}}]{Johnson2009}%
	\BibitemOpen
	\bibfield  {author} {\bibinfo {author} {\bibfnamefont {P.~R.}\ \bibnamefont
			{Johnson}}, \bibinfo {author} {\bibfnamefont {E.}~\bibnamefont {Tiesinga}},
		\bibinfo {author} {\bibfnamefont {J.~V.}\ \bibnamefont {Porto}},\ and\
		\bibinfo {author} {\bibfnamefont {C.~J.}\ \bibnamefont {Williams}},\
	}\bibfield  {title} {\bibinfo {title} {Effective three-body interactions of
			neutral bosons in optical lattices},\ }\href
	{https://doi.org/10.1088/1367-2630/11/9/093022} {\bibfield  {journal}
		{\bibinfo  {journal} {New J. Phys.}\ }\textbf {\bibinfo {volume} {11}},\
		\bibinfo {pages} {093022} (\bibinfo {year} {2009})}\BibitemShut {NoStop}%
	\bibitem [{\citenamefont {Tiesinga}\ and\ \citenamefont
		{Johnson}(2011)}]{Tiesinga2011}%
	\BibitemOpen
	\bibfield  {author} {\bibinfo {author} {\bibfnamefont {E.}~\bibnamefont
			{Tiesinga}}\ and\ \bibinfo {author} {\bibfnamefont {P.~R.}\ \bibnamefont
			{Johnson}},\ }\bibfield  {title} {\bibinfo {title} {Collapse and revival
			dynamics of number-squeezed superfluids of ultracold atoms in optical
			lattices},\ }\href {https://doi.org/10.1103/PhysRevA.83.063609} {\bibfield
		{journal} {\bibinfo  {journal} {Phys. Rev. A}\ }\textbf {\bibinfo {volume}
			{83}},\ \bibinfo {pages} {063609} (\bibinfo {year} {2011})}\BibitemShut
	{NoStop}%
	\bibitem [{\citenamefont {Johnson}\ \emph {et~al.}(2012)\citenamefont
		{Johnson}, \citenamefont {Blume}, \citenamefont {Yin}, \citenamefont
		{Flynn},\ and\ \citenamefont {Tiesinga}}]{Johnson_2012}%
	\BibitemOpen
	\bibfield  {author} {\bibinfo {author} {\bibfnamefont {P.~R.}\ \bibnamefont
			{Johnson}}, \bibinfo {author} {\bibfnamefont {D.}~\bibnamefont {Blume}},
		\bibinfo {author} {\bibfnamefont {X.~Y.}\ \bibnamefont {Yin}}, \bibinfo
		{author} {\bibfnamefont {W.~F.}\ \bibnamefont {Flynn}},\ and\ \bibinfo
		{author} {\bibfnamefont {E.}~\bibnamefont {Tiesinga}},\ }\bibfield  {title}
	{\bibinfo {title} {Effective renormalized multi-body interactions of
			harmonically confined ultracold neutral bosons},\ }\href
	{https://doi.org/10.1088/1367-2630/14/5/053037} {\bibfield  {journal}
		{\bibinfo  {journal} {New J. Phys.}\ }\textbf {\bibinfo {volume} {14}},\
		\bibinfo {pages} {053037} (\bibinfo {year} {2012})}\BibitemShut {NoStop}%
	\bibitem [{\citenamefont {Yin}\ \emph {et~al.}(2014)\citenamefont {Yin},
		\citenamefont {Blume}, \citenamefont {Johnson},\ and\ \citenamefont
		{Tiesinga}}]{PhysRevA.90.043631}%
	\BibitemOpen
	\bibfield  {author} {\bibinfo {author} {\bibfnamefont {X.~Y.}\ \bibnamefont
			{Yin}}, \bibinfo {author} {\bibfnamefont {D.}~\bibnamefont {Blume}}, \bibinfo
		{author} {\bibfnamefont {P.~R.}\ \bibnamefont {Johnson}},\ and\ \bibinfo
		{author} {\bibfnamefont {E.}~\bibnamefont {Tiesinga}},\ }\bibfield  {title}
	{\bibinfo {title} {Universal and nonuniversal effective ${N}$-body
			interactions for ultracold harmonically trapped few-atom systems},\ }\href
	{https://doi.org/10.1103/PhysRevA.90.043631} {\bibfield  {journal} {\bibinfo
			{journal} {Phys. Rev. A}\ }\textbf {\bibinfo {volume} {90}},\ \bibinfo
		{pages} {043631} (\bibinfo {year} {2014})}\BibitemShut {NoStop}%
	\bibitem [{\citenamefont {Perlin}\ and\ \citenamefont
		{Rey}(2019)}]{Perlin_2019}%
	\BibitemOpen
	\bibfield  {author} {\bibinfo {author} {\bibfnamefont {M.~A.}\ \bibnamefont
			{Perlin}}\ and\ \bibinfo {author} {\bibfnamefont {A.~M.}\ \bibnamefont
			{Rey}},\ }\bibfield  {title} {\bibinfo {title} {Effective multi-body
			{SU}(${N}$)-symmetric interactions of ultracold fermionic atoms on a 3{D}
			lattice},\ }\href {https://doi.org/10.1088/1367-2630/ab0e50} {\bibfield
		{journal} {\bibinfo  {journal} {New J. Phys.}\ }\textbf {\bibinfo {volume}
			{21}},\ \bibinfo {pages} {043039} (\bibinfo {year} {2019})}\BibitemShut
	{NoStop}%
	\bibitem [{\citenamefont {Mahmud}\ \emph {et~al.}(2014)\citenamefont {Mahmud},
		\citenamefont {Tiesinga},\ and\ \citenamefont
		{Johnson}}]{PhysRevA.90.041602}%
	\BibitemOpen
	\bibfield  {author} {\bibinfo {author} {\bibfnamefont {K.~W.}\ \bibnamefont
			{Mahmud}}, \bibinfo {author} {\bibfnamefont {E.}~\bibnamefont {Tiesinga}},\
		and\ \bibinfo {author} {\bibfnamefont {P.~R.}\ \bibnamefont {Johnson}},\
	}\bibfield  {title} {\bibinfo {title} {Dynamically decoupled three-body
			interactions with applications to interaction-based quantum metrology},\
	}\href {https://doi.org/10.1103/PhysRevA.90.041602} {\bibfield  {journal}
		{\bibinfo  {journal} {Phys. Rev. A}\ }\textbf {\bibinfo {volume} {90}},\
		\bibinfo {pages} {041602(R)} (\bibinfo {year} {2014})}\BibitemShut {NoStop}%
	\bibitem [{\citenamefont {Will}\ \emph {et~al.}(2010)\citenamefont {Will},
		\citenamefont {Best}, \citenamefont {Schneider}, \citenamefont
		{Hackerm{\"u}ller}, \citenamefont {L{\"u}hmann},\ and\ \citenamefont
		{Bloch}}]{Will2010}%
	\BibitemOpen
	\bibfield  {author} {\bibinfo {author} {\bibfnamefont {S.}~\bibnamefont
			{Will}}, \bibinfo {author} {\bibfnamefont {T.}~\bibnamefont {Best}}, \bibinfo
		{author} {\bibfnamefont {U.}~\bibnamefont {Schneider}}, \bibinfo {author}
		{\bibfnamefont {L.}~\bibnamefont {Hackerm{\"u}ller}}, \bibinfo {author}
		{\bibfnamefont {D.-S.}\ \bibnamefont {L{\"u}hmann}},\ and\ \bibinfo {author}
		{\bibfnamefont {I.}~\bibnamefont {Bloch}},\ }\bibfield  {title} {\bibinfo
		{title} {Time-resolved observation of coherent multi-body interactions in
			quantum phase revivals},\ }\href {https://doi.org/10.1038/nature09036}
	{\bibfield  {journal} {\bibinfo  {journal} {Nature}\ }\textbf {\bibinfo
			{volume} {465}},\ \bibinfo {pages} {197–201} (\bibinfo {year}
		{2010})}\BibitemShut {NoStop}%
	\bibitem [{\citenamefont {Ma}\ \emph {et~al.}(2011)\citenamefont {Ma},
		\citenamefont {Tai}, \citenamefont {Preiss}, \citenamefont {Bakr},
		\citenamefont {Simon},\ and\ \citenamefont {Greiner}}]{ma2011photon}%
	\BibitemOpen
	\bibfield  {author} {\bibinfo {author} {\bibfnamefont {R.}~\bibnamefont
			{Ma}}, \bibinfo {author} {\bibfnamefont {M.~E.}\ \bibnamefont {Tai}},
		\bibinfo {author} {\bibfnamefont {P.~M.}\ \bibnamefont {Preiss}}, \bibinfo
		{author} {\bibfnamefont {W.~S.}\ \bibnamefont {Bakr}}, \bibinfo {author}
		{\bibfnamefont {J.}~\bibnamefont {Simon}},\ and\ \bibinfo {author}
		{\bibfnamefont {M.}~\bibnamefont {Greiner}},\ }\bibfield  {title} {\bibinfo
		{title} {Photon-assisted tunneling in a biased strongly correlated {B}ose
			gas},\ }\href {https://doi.org/10.1103/PhysRevLett.107.095301} {\bibfield
		{journal} {\bibinfo  {journal} {Phys. Rev. Lett.}\ }\textbf {\bibinfo
			{volume} {107}},\ \bibinfo {pages} {095301} (\bibinfo {year}
		{2011})}\BibitemShut {NoStop}%
	\bibitem [{\citenamefont {Mark}\ \emph {et~al.}(2011)\citenamefont {Mark},
		\citenamefont {Haller}, \citenamefont {Lauber}, \citenamefont {Danzl},
		\citenamefont {Daley},\ and\ \citenamefont {N{\"a}gerl}}]{mark2011precision}%
	\BibitemOpen
	\bibfield  {author} {\bibinfo {author} {\bibfnamefont {M.~J.}\ \bibnamefont
			{Mark}}, \bibinfo {author} {\bibfnamefont {E.}~\bibnamefont {Haller}},
		\bibinfo {author} {\bibfnamefont {K.}~\bibnamefont {Lauber}}, \bibinfo
		{author} {\bibfnamefont {J.~G.}\ \bibnamefont {Danzl}}, \bibinfo {author}
		{\bibfnamefont {A.}~\bibnamefont {Daley}},\ and\ \bibinfo {author}
		{\bibfnamefont {H.-C.}\ \bibnamefont {N{\"a}gerl}},\ }\bibfield  {title}
	{\bibinfo {title} {Precision measurements on a tunable {M}ott insulator of
			ultracold atoms},\ }\href {https://doi.org/10.1103/PhysRevLett.107.175301}
	{\bibfield  {journal} {\bibinfo  {journal} {Phys. Rev. Lett.}\ }\textbf
		{\bibinfo {volume} {107}},\ \bibinfo {pages} {175301} (\bibinfo {year}
		{2011})}\BibitemShut {NoStop}%
	\bibitem [{\citenamefont {Goban}\ \emph {et~al.}(2018)\citenamefont {Goban},
		\citenamefont {Hutson}, \citenamefont {Marti}, \citenamefont {Campbell},
		\citenamefont {Perlin}, \citenamefont {Julienne}, \citenamefont {D'Incao},
		\citenamefont {Ray},\ and\ \citenamefont {Ye}}]{Goban2018Emergence}%
	\BibitemOpen
	\bibfield  {author} {\bibinfo {author} {\bibfnamefont {A.}~\bibnamefont
			{Goban}}, \bibinfo {author} {\bibfnamefont {R.~B.}\ \bibnamefont {Hutson}},
		\bibinfo {author} {\bibfnamefont {G.~E.}\ \bibnamefont {Marti}}, \bibinfo
		{author} {\bibfnamefont {S.~L.}\ \bibnamefont {Campbell}}, \bibinfo {author}
		{\bibfnamefont {M.~A.}\ \bibnamefont {Perlin}}, \bibinfo {author}
		{\bibfnamefont {P.~S.}\ \bibnamefont {Julienne}}, \bibinfo {author}
		{\bibfnamefont {J.~P.}\ \bibnamefont {D'Incao}}, \bibinfo {author}
		{\bibfnamefont {A.~M.}\ \bibnamefont {Ray}},\ and\ \bibinfo {author}
		{\bibfnamefont {J.}~\bibnamefont {Ye}},\ }\bibfield  {title} {\bibinfo
		{title} {Emergence of multi-body interactions in a fermionic lattice clock},\
	}\href {https://doi.org/10.1038/s41586-018-0661-6} {\bibfield  {journal}
		{\bibinfo  {journal} {Nature}\ }\textbf {\bibinfo {volume} {563}},\ \bibinfo
		{pages} {369–373} (\bibinfo {year} {2018})}\BibitemShut {NoStop}%
	\bibitem [{\citenamefont {Gambetta}\ \emph
		{et~al.}(2020{\natexlab{a}})\citenamefont {Gambetta}, \citenamefont {Li},
		\citenamefont {Schmidt-Kaler},\ and\ \citenamefont
		{Lesanovsky}}]{Gambetta2020Engineering}%
	\BibitemOpen
	\bibfield  {author} {\bibinfo {author} {\bibfnamefont {F.~M.}\ \bibnamefont
			{Gambetta}}, \bibinfo {author} {\bibfnamefont {W.}~\bibnamefont {Li}},
		\bibinfo {author} {\bibfnamefont {F.}~\bibnamefont {Schmidt-Kaler}},\ and\
		\bibinfo {author} {\bibfnamefont {I.}~\bibnamefont {Lesanovsky}},\ }\bibfield
	{title} {\bibinfo {title} {Engineering nonbinary {R}ydberg interactions via
			phonons in an optical lattice},\ }\href
	{https://doi.org/10.1103/PhysRevLett.124.043402} {\bibfield  {journal}
		{\bibinfo  {journal} {Phys. Rev. Lett.}\ }\textbf {\bibinfo {volume} {124}},\
		\bibinfo {pages} {043402} (\bibinfo {year} {2020}{\natexlab{a}})}\BibitemShut
	{NoStop}%
	\bibitem [{\citenamefont {Gambetta}\ \emph
		{et~al.}(2020{\natexlab{b}})\citenamefont {Gambetta}, \citenamefont {Zhang},
		\citenamefont {Hennrich}, \citenamefont {Lesanovsky},\ and\ \citenamefont
		{Li}}]{Gambetta2020LongRange}%
	\BibitemOpen
	\bibfield  {author} {\bibinfo {author} {\bibfnamefont {F.~M.}\ \bibnamefont
			{Gambetta}}, \bibinfo {author} {\bibfnamefont {C.}~\bibnamefont {Zhang}},
		\bibinfo {author} {\bibfnamefont {M.}~\bibnamefont {Hennrich}}, \bibinfo
		{author} {\bibfnamefont {I.}~\bibnamefont {Lesanovsky}},\ and\ \bibinfo
		{author} {\bibfnamefont {W.}~\bibnamefont {Li}},\ }\bibfield  {title}
	{\bibinfo {title} {Long-range multibody interactions and three-body
			antiblockade in a trapped {R}ydberg ion chain},\ }\href
	{https://doi.org/10.1103/PhysRevLett.125.133602} {\bibfield  {journal}
		{\bibinfo  {journal} {Phys. Rev. Lett.}\ }\textbf {\bibinfo {volume} {125}},\
		\bibinfo {pages} {133602} (\bibinfo {year} {2020}{\natexlab{b}})}\BibitemShut
	{NoStop}%
	\bibitem [{\citenamefont {Fey}\ \emph {et~al.}(2019)\citenamefont {Fey},
		\citenamefont {Yang}, \citenamefont {Rittenhouse}, \citenamefont {Munkes},
		\citenamefont {Baluktsian}, \citenamefont {Schmelcher}, \citenamefont
		{Sadeghpour},\ and\ \citenamefont {Shaffer}}]{Fey2019Effective}%
	\BibitemOpen
	\bibfield  {author} {\bibinfo {author} {\bibfnamefont {C.}~\bibnamefont
			{Fey}}, \bibinfo {author} {\bibfnamefont {J.}~\bibnamefont {Yang}}, \bibinfo
		{author} {\bibfnamefont {S.~T.}\ \bibnamefont {Rittenhouse}}, \bibinfo
		{author} {\bibfnamefont {F.}~\bibnamefont {Munkes}}, \bibinfo {author}
		{\bibfnamefont {M.}~\bibnamefont {Baluktsian}}, \bibinfo {author}
		{\bibfnamefont {P.}~\bibnamefont {Schmelcher}}, \bibinfo {author}
		{\bibfnamefont {H.~R.}\ \bibnamefont {Sadeghpour}},\ and\ \bibinfo {author}
		{\bibfnamefont {J.~P.}\ \bibnamefont {Shaffer}},\ }\bibfield  {title}
	{\bibinfo {title} {Effective three-body interactions in
			$\mathrm{Cs}(6s)\text{\ensuremath{-}}\mathrm{Cs}(nd)$ {R}ydberg trimers},\
	}\href {https://doi.org/10.1103/PhysRevLett.122.103001} {\bibfield  {journal}
		{\bibinfo  {journal} {Phys. Rev. Lett.}\ }\textbf {\bibinfo {volume} {122}},\
		\bibinfo {pages} {103001} (\bibinfo {year} {2019})}\BibitemShut {NoStop}%
	\bibitem [{\citenamefont {Schweizer}\ \emph {et~al.}(2019)\citenamefont
		{Schweizer}, \citenamefont {Grusdt}, \citenamefont {Berngruber},
		\citenamefont {Barbiero}, \citenamefont {Demler}, \citenamefont {Goldman},
		\citenamefont {Bloch},\ and\ \citenamefont {Aidelsburger}}]{Schweizer_2019}%
	\BibitemOpen
	\bibfield  {author} {\bibinfo {author} {\bibfnamefont {C.}~\bibnamefont
			{Schweizer}}, \bibinfo {author} {\bibfnamefont {F.}~\bibnamefont {Grusdt}},
		\bibinfo {author} {\bibfnamefont {M.}~\bibnamefont {Berngruber}}, \bibinfo
		{author} {\bibfnamefont {L.}~\bibnamefont {Barbiero}}, \bibinfo {author}
		{\bibfnamefont {E.}~\bibnamefont {Demler}}, \bibinfo {author} {\bibfnamefont
			{N.}~\bibnamefont {Goldman}}, \bibinfo {author} {\bibfnamefont
			{I.}~\bibnamefont {Bloch}},\ and\ \bibinfo {author} {\bibfnamefont
			{M.}~\bibnamefont {Aidelsburger}},\ }\bibfield  {title} {\bibinfo {title}
		{Floquet approach to $\mathbb{Z}_2$ lattice gauge theories with ultracold
			atoms in optical lattices},\ }\href
	{https://doi.org/10.1038/s41567-019-0649-7} {\bibfield  {journal} {\bibinfo
			{journal} {Nature Phys.}\ }\textbf {\bibinfo {volume} {15}},\ \bibinfo
		{pages} {1168–1173} (\bibinfo {year} {2019})}\BibitemShut {NoStop}%
	\bibitem [{\citenamefont {Bañuls}\ \emph {et~al.}(2020)\citenamefont
		{Bañuls}, \citenamefont {Blatt}, \citenamefont {Catani}, \citenamefont
		{Celi}, \citenamefont {Cirac}, \citenamefont {Dalmonte}, \citenamefont
		{Fallani}, \citenamefont {Jansen}, \citenamefont {Lewenstein}, \citenamefont
		{Montangero}, \citenamefont {Muschik}, \citenamefont {Reznik}, \citenamefont
		{Rico}, \citenamefont {Tagliacozzo}, \citenamefont {Van~Acoleyen},
		\citenamefont {Verstraete}, \citenamefont {Wiese}, \citenamefont {Wingate},
		\citenamefont {Zakrzewski},\ and\ \citenamefont {Zoller}}]{Banuls_2020}%
	\BibitemOpen
	\bibfield  {author} {\bibinfo {author} {\bibfnamefont {M.~C.}\ \bibnamefont
			{Bañuls}}, \bibinfo {author} {\bibfnamefont {R.}~\bibnamefont {Blatt}},
		\bibinfo {author} {\bibfnamefont {J.}~\bibnamefont {Catani}}, \bibinfo
		{author} {\bibfnamefont {A.}~\bibnamefont {Celi}}, \bibinfo {author}
		{\bibfnamefont {J.~I.}\ \bibnamefont {Cirac}}, \bibinfo {author}
		{\bibfnamefont {M.}~\bibnamefont {Dalmonte}}, \bibinfo {author}
		{\bibfnamefont {L.}~\bibnamefont {Fallani}}, \bibinfo {author} {\bibfnamefont
			{K.}~\bibnamefont {Jansen}}, \bibinfo {author} {\bibfnamefont
			{M.}~\bibnamefont {Lewenstein}}, \bibinfo {author} {\bibfnamefont
			{S.}~\bibnamefont {Montangero}}, \bibinfo {author} {\bibfnamefont {C.~A.}\
			\bibnamefont {Muschik}}, \bibinfo {author} {\bibfnamefont {B.}~\bibnamefont
			{Reznik}}, \bibinfo {author} {\bibfnamefont {E.}~\bibnamefont {Rico}},
		\bibinfo {author} {\bibfnamefont {L.}~\bibnamefont {Tagliacozzo}}, \bibinfo
		{author} {\bibfnamefont {K.}~\bibnamefont {Van~Acoleyen}}, \bibinfo {author}
		{\bibfnamefont {F.}~\bibnamefont {Verstraete}}, \bibinfo {author}
		{\bibfnamefont {U.-J.}\ \bibnamefont {Wiese}}, \bibinfo {author}
		{\bibfnamefont {M.}~\bibnamefont {Wingate}}, \bibinfo {author} {\bibfnamefont
			{J.}~\bibnamefont {Zakrzewski}},\ and\ \bibinfo {author} {\bibfnamefont
			{P.}~\bibnamefont {Zoller}},\ }\bibfield  {title} {\bibinfo {title}
		{Simulating lattice gauge theories within quantum technologies},\ }\href
	{https://doi.org/10.1140/epjd/e2020-100571-8} {\bibfield  {journal} {\bibinfo
			{journal} {Euro. Phys. J. D}\ }\textbf {\bibinfo {volume} {74}},\ \bibinfo
		{pages} {165} (\bibinfo {year} {2020})}\BibitemShut {NoStop}%
	\bibitem [{\citenamefont {Hauke}\ \emph {et~al.}(2013)\citenamefont {Hauke},
		\citenamefont {Marcos}, \citenamefont {Dalmonte},\ and\ \citenamefont
		{Zoller}}]{PhysRevX.3.041018}%
	\BibitemOpen
	\bibfield  {author} {\bibinfo {author} {\bibfnamefont {P.}~\bibnamefont
			{Hauke}}, \bibinfo {author} {\bibfnamefont {D.}~\bibnamefont {Marcos}},
		\bibinfo {author} {\bibfnamefont {M.}~\bibnamefont {Dalmonte}},\ and\
		\bibinfo {author} {\bibfnamefont {P.}~\bibnamefont {Zoller}},\ }\bibfield
	{title} {\bibinfo {title} {Quantum simulation of a lattice {S}chwinger model
			in a chain of trapped ions},\ }\href
	{https://doi.org/10.1103/PhysRevX.3.041018} {\bibfield  {journal} {\bibinfo
			{journal} {Phys. Rev. X}\ }\textbf {\bibinfo {volume} {3}},\ \bibinfo {pages}
		{041018} (\bibinfo {year} {2013})}\BibitemShut {NoStop}%
	\bibitem [{\citenamefont {Zohar}\ \emph {et~al.}(2017)\citenamefont {Zohar},
		\citenamefont {Farace}, \citenamefont {Reznik},\ and\ \citenamefont
		{Cirac}}]{PhysRevLett.118.070501}%
	\BibitemOpen
	\bibfield  {author} {\bibinfo {author} {\bibfnamefont {E.}~\bibnamefont
			{Zohar}}, \bibinfo {author} {\bibfnamefont {A.}~\bibnamefont {Farace}},
		\bibinfo {author} {\bibfnamefont {B.}~\bibnamefont {Reznik}},\ and\ \bibinfo
		{author} {\bibfnamefont {J.~I.}\ \bibnamefont {Cirac}},\ }\bibfield  {title}
	{\bibinfo {title} {Digital quantum simulation of $\mathbb{Z}_{2}$ lattice
			gauge theories with dynamical fermionic matter},\ }\href
	{https://doi.org/10.1103/PhysRevLett.118.070501} {\bibfield  {journal}
		{\bibinfo  {journal} {Phys. Rev. Lett.}\ }\textbf {\bibinfo {volume} {118}},\
		\bibinfo {pages} {070501} (\bibinfo {year} {2017})}\BibitemShut {NoStop}%
	\bibitem [{\citenamefont {Motrunich}(2005)}]{Motrunich2005Variational}%
	\BibitemOpen
	\bibfield  {author} {\bibinfo {author} {\bibfnamefont {O.~I.}\ \bibnamefont
			{Motrunich}},\ }\bibfield  {title} {\bibinfo {title} {Variational study of
			triangular lattice spin-$1/2$ model with ring exchanges and spin liquid state
			in
			$\ensuremath{\kappa}\text{\ensuremath{-}}{(\mathrm{ET})}_{2}\mathrm{{C}u}_{2}{(\mathrm{CN})}_{3}$},\
	}\href {https://doi.org/10.1103/PhysRevB.72.045105} {\bibfield  {journal}
		{\bibinfo  {journal} {Phys. Rev. B}\ }\textbf {\bibinfo {volume} {72}},\
		\bibinfo {pages} {045105} (\bibinfo {year} {2005})}\BibitemShut {NoStop}%
	\bibitem [{\citenamefont {Andrade}\ \emph {et~al.}(2022)\citenamefont
		{Andrade}, \citenamefont {Davoudi}, \citenamefont {Graß}, \citenamefont
		{Hafezi}, \citenamefont {Pagano},\ and\ \citenamefont {Seif}}]{Andrade2022}%
	\BibitemOpen
	\bibfield  {author} {\bibinfo {author} {\bibfnamefont {B.}~\bibnamefont
			{Andrade}}, \bibinfo {author} {\bibfnamefont {Z.}~\bibnamefont {Davoudi}},
		\bibinfo {author} {\bibfnamefont {T.}~\bibnamefont {Graß}}, \bibinfo
		{author} {\bibfnamefont {M.}~\bibnamefont {Hafezi}}, \bibinfo {author}
		{\bibfnamefont {G.}~\bibnamefont {Pagano}},\ and\ \bibinfo {author}
		{\bibfnamefont {A.}~\bibnamefont {Seif}},\ }\bibfield  {title} {\bibinfo
		{title} {Engineering an effective three-spin {H}amiltonian in trapped-ion
			systems for applications in quantum simulation},\ }\href
	{https://doi.org/10.1088/2058-9565/ac5f5b} {\bibfield  {journal} {\bibinfo
			{journal} {Quantum Sci. Technol.}\ }\textbf {\bibinfo {volume} {7}},\
		\bibinfo {pages} {034001} (\bibinfo {year} {2022})}\BibitemShut {NoStop}%
	\bibitem [{\citenamefont {Pachos}\ and\ \citenamefont
		{Plenio}(2004)}]{PhysRevLett.93.056402}%
	\BibitemOpen
	\bibfield  {author} {\bibinfo {author} {\bibfnamefont {J.~K.}\ \bibnamefont
			{Pachos}}\ and\ \bibinfo {author} {\bibfnamefont {M.~B.}\ \bibnamefont
			{Plenio}},\ }\bibfield  {title} {\bibinfo {title} {Three-spin interactions in
			optical lattices and criticality in cluster {H}amiltonians},\ }\href
	{https://doi.org/10.1103/PhysRevLett.93.056402} {\bibfield  {journal}
		{\bibinfo  {journal} {Phys. Rev. Lett.}\ }\textbf {\bibinfo {volume} {93}},\
		\bibinfo {pages} {056402} (\bibinfo {year} {2004})}\BibitemShut {NoStop}%
	\bibitem [{\citenamefont {Büchler}\ \emph {et~al.}(2007)\citenamefont
		{Büchler}, \citenamefont {Micheli},\ and\ \citenamefont
		{Zoller}}]{Buchler_2007}%
	\BibitemOpen
	\bibfield  {author} {\bibinfo {author} {\bibfnamefont {H.~P.}\ \bibnamefont
			{Büchler}}, \bibinfo {author} {\bibfnamefont {A.}~\bibnamefont {Micheli}},\
		and\ \bibinfo {author} {\bibfnamefont {P.}~\bibnamefont {Zoller}},\
	}\bibfield  {title} {\bibinfo {title} {Three-body interactions with cold
			polar molecules},\ }\href {https://doi.org/10.1038/nphys678} {\bibfield
		{journal} {\bibinfo  {journal} {Nature Phys.}\ }\textbf {\bibinfo {volume}
			{3}},\ \bibinfo {pages} {726–731} (\bibinfo {year} {2007})}\BibitemShut
	{NoStop}%
	\bibitem [{\citenamefont {Schmidt}\ \emph {et~al.}(2008)\citenamefont
		{Schmidt}, \citenamefont {Dorier},\ and\ \citenamefont
		{L\"auchli}}]{PhysRevLett.101.150405}%
	\BibitemOpen
	\bibfield  {author} {\bibinfo {author} {\bibfnamefont {K.~P.}\ \bibnamefont
			{Schmidt}}, \bibinfo {author} {\bibfnamefont {J.}~\bibnamefont {Dorier}},\
		and\ \bibinfo {author} {\bibfnamefont {A.~M.}\ \bibnamefont {L\"auchli}},\
	}\bibfield  {title} {\bibinfo {title} {Solids and supersolids of three-body
			interacting polar molecules on an optical lattice},\ }\href
	{https://doi.org/10.1103/PhysRevLett.101.150405} {\bibfield  {journal}
		{\bibinfo  {journal} {Phys. Rev. Lett.}\ }\textbf {\bibinfo {volume} {101}},\
		\bibinfo {pages} {150405} (\bibinfo {year} {2008})}\BibitemShut {NoStop}%
	\bibitem [{\citenamefont {Harshman}\ and\ \citenamefont
		{Knapp}(2020)}]{Harshman_2020}%
	\BibitemOpen
	\bibfield  {author} {\bibinfo {author} {\bibfnamefont {N.~L.}\ \bibnamefont
			{Harshman}}\ and\ \bibinfo {author} {\bibfnamefont {A.~C.}\ \bibnamefont
			{Knapp}},\ }\bibfield  {title} {\bibinfo {title} {Anyons from three-body
			hard-core interactions in one dimension},\ }\href
	{https://doi.org/10.1016/j.aop.2019.168003} {\bibfield  {journal} {\bibinfo
			{journal} {Ann. Phys.}\ }\textbf {\bibinfo {volume} {412}},\ \bibinfo {pages}
		{168003} (\bibinfo {year} {2020})}\BibitemShut {NoStop}%
	\bibitem [{\citenamefont {Singh}\ and\ \citenamefont
		{Mishra}(2016)}]{PhysRevA.94.063610}%
	\BibitemOpen
	\bibfield  {author} {\bibinfo {author} {\bibfnamefont {M.}~\bibnamefont
			{Singh}}\ and\ \bibinfo {author} {\bibfnamefont {T.}~\bibnamefont {Mishra}},\
	}\bibfield  {title} {\bibinfo {title} {Three-body interacting dipolar bosons
			and the fate of lattice supersolidity},\ }\href
	{https://doi.org/10.1103/PhysRevA.94.063610} {\bibfield  {journal} {\bibinfo
			{journal} {Phys. Rev. A}\ }\textbf {\bibinfo {volume} {94}},\ \bibinfo
		{pages} {063610} (\bibinfo {year} {2016})}\BibitemShut {NoStop}%
	\bibitem [{\citenamefont {Honda}\ \emph {et~al.}(2025)\citenamefont {Honda},
		\citenamefont {Takasu}, \citenamefont {Haruna}, \citenamefont {Nishida},\
		and\ \citenamefont {Takahashi}}]{Honda2025Exploring}%
	\BibitemOpen
	\bibfield  {author} {\bibinfo {author} {\bibfnamefont {K.}~\bibnamefont
			{Honda}}, \bibinfo {author} {\bibfnamefont {Y.}~\bibnamefont {Takasu}},
		\bibinfo {author} {\bibfnamefont {Y.}~\bibnamefont {Haruna}}, \bibinfo
		{author} {\bibfnamefont {Y.}~\bibnamefont {Nishida}},\ and\ \bibinfo {author}
		{\bibfnamefont {Y.}~\bibnamefont {Takahashi}},\ }\bibfield  {title} {\bibinfo
		{title} {Exploring the strongly interacting regime of effective multibody
			interactions in a trapped ultracold-atom system},\ }\href
	{https://doi.org/10.1103/PhysRevA.111.033303} {\bibfield  {journal} {\bibinfo
			{journal} {Phys. Rev. A.}\ }\textbf {\bibinfo {volume} {111}},\ \bibinfo
		{pages} {033303} (\bibinfo {year} {2025})}\BibitemShut {NoStop}%
	\bibitem [{\citenamefont {Endo}\ \emph {et~al.}(2025)\citenamefont {Endo},
		\citenamefont {Epelbaum}, \citenamefont {Naidon}, \citenamefont {Nishida},
		\citenamefont {Sekiguchi},\ and\ \citenamefont
		{Takahashi}}]{Endo2025ThreeBody}%
	\BibitemOpen
	\bibfield  {author} {\bibinfo {author} {\bibfnamefont {S.}~\bibnamefont
			{Endo}}, \bibinfo {author} {\bibfnamefont {E.}~\bibnamefont {Epelbaum}},
		\bibinfo {author} {\bibfnamefont {P.}~\bibnamefont {Naidon}}, \bibinfo
		{author} {\bibfnamefont {Y.}~\bibnamefont {Nishida}}, \bibinfo {author}
		{\bibfnamefont {K.}~\bibnamefont {Sekiguchi}},\ and\ \bibinfo {author}
		{\bibfnamefont {Y.}~\bibnamefont {Takahashi}},\ }\bibfield  {title} {\bibinfo
		{title} {Three-body forces and \uppercase{E}fimov physics in nuclei and
			atoms},\ }\href {https://doi.org/10.1140/epja/s10050-024-01467-4} {\bibfield
		{journal} {\bibinfo  {journal} {Eur. Phys. J. A}\ }\textbf {\bibinfo {volume}
			{61}},\ \bibinfo {pages} {9} (\bibinfo {year} {2025})}\BibitemShut {NoStop}%
	\bibitem [{\citenamefont {Braaten}\ and\ \citenamefont
		{Hammer}(2007)}]{Braaten2007Efimov}%
	\BibitemOpen
	\bibfield  {author} {\bibinfo {author} {\bibfnamefont {E.}~\bibnamefont
			{Braaten}}\ and\ \bibinfo {author} {\bibfnamefont {H.-W.}\ \bibnamefont
			{Hammer}},\ }\bibfield  {title} {\bibinfo {title} {{E}fimov physics in cold
			atoms},\ }\href {https://doi.org/https://doi.org/10.1016/j.aop.2006.10.011}
	{\bibfield  {journal} {\bibinfo  {journal} {Ann. Phys.}\ }\textbf {\bibinfo
			{volume} {322}},\ \bibinfo {pages} {120–163} (\bibinfo {year}
		{2007})}\BibitemShut {NoStop}%
	\bibitem [{\citenamefont {Secker}\ \emph {et~al.}(2021)\citenamefont {Secker},
		\citenamefont {Li}, \citenamefont {Mestrom},\ and\ \citenamefont
		{Kokkelmans}}]{Secker2021Multichannel}%
	\BibitemOpen
	\bibfield  {author} {\bibinfo {author} {\bibfnamefont {T.}~\bibnamefont
			{Secker}}, \bibinfo {author} {\bibfnamefont {J.-L.}\ \bibnamefont {Li}},
		\bibinfo {author} {\bibfnamefont {P.~M.~A.}\ \bibnamefont {Mestrom}},\ and\
		\bibinfo {author} {\bibfnamefont {S.~J. J. M.~F.}\ \bibnamefont
			{Kokkelmans}},\ }\bibfield  {title} {\bibinfo {title} {Multichannel nature of
			three-body recombination for ultracold $^{39}\mathrm{K}$},\ }\href
	{https://doi.org/10.1103/PhysRevA.103.022825} {\bibfield  {journal} {\bibinfo
			{journal} {Phys. Rev. A}\ }\textbf {\bibinfo {volume} {103}},\ \bibinfo
		{pages} {022825} (\bibinfo {year} {2021})}\BibitemShut {NoStop}%
	\bibitem [{\citenamefont {Musolino}\ \emph {et~al.}(2022)\citenamefont
		{Musolino}, \citenamefont {Kurkjian}, \citenamefont {Van~Regemortel},
		\citenamefont {Wouters}, \citenamefont {Kokkelmans},\ and\ \citenamefont
		{Colussi}}]{Musolino2022}%
	\BibitemOpen
	\bibfield  {author} {\bibinfo {author} {\bibfnamefont {S.}~\bibnamefont
			{Musolino}}, \bibinfo {author} {\bibfnamefont {H.}~\bibnamefont {Kurkjian}},
		\bibinfo {author} {\bibfnamefont {M.}~\bibnamefont {Van~Regemortel}},
		\bibinfo {author} {\bibfnamefont {M.}~\bibnamefont {Wouters}}, \bibinfo
		{author} {\bibfnamefont {S.~J. J. M.~F.}\ \bibnamefont {Kokkelmans}},\ and\
		\bibinfo {author} {\bibfnamefont {V.~E.}\ \bibnamefont {Colussi}},\
	}\bibfield  {title} {\bibinfo {title} {Bose-{E}instein condensation of
			{E}fimovian triples in the unitary {B}ose gas},\ }\href
	{https://journals.aps.org/prl/abstract/10.1103/PhysRevLett.128.020401}
	{\bibfield  {journal} {\bibinfo  {journal} {Phys. Rev. Lett.}\ }\textbf
		{\bibinfo {volume} {128}},\ \bibinfo {pages} {020401} (\bibinfo {year}
		{2022})}\BibitemShut {NoStop}%
	\bibitem [{\citenamefont {van~de Kraats}\ \emph {et~al.}(2024)\citenamefont
		{van~de Kraats}, \citenamefont {Ahmed-Braun}, \citenamefont {Li},\ and\
		\citenamefont {Kokkelmans}}]{Kraats2024Emergent}%
	\BibitemOpen
	\bibfield  {author} {\bibinfo {author} {\bibfnamefont {J.}~\bibnamefont
			{van~de Kraats}}, \bibinfo {author} {\bibfnamefont {D.~J.~M.}\ \bibnamefont
			{Ahmed-Braun}}, \bibinfo {author} {\bibfnamefont {J.-L.}\ \bibnamefont
			{Li}},\ and\ \bibinfo {author} {\bibfnamefont {S.~J. J. M.~F.}\ \bibnamefont
			{Kokkelmans}},\ }\bibfield  {title} {\bibinfo {title} {Emergent inflation of
			the {E}fimov spectrum under three-body spin-exchange interactions},\ }\href
	{https://doi.org/10.1103/PhysRevLett.132.133402} {\bibfield  {journal}
		{\bibinfo  {journal} {Phys. Rev. Lett.}\ }\textbf {\bibinfo {volume} {132}},\
		\bibinfo {pages} {133402} (\bibinfo {year} {2024})}\BibitemShut {NoStop}%
	\bibitem [{\citenamefont {Cieśliński}\ \emph {et~al.}(2023)\citenamefont
		{Cieśliński}, \citenamefont {Kłobus}, \citenamefont {Kurzyński},
		\citenamefont {Paterek},\ and\ \citenamefont {Laskowski}}]{Cieslinski_2023}%
	\BibitemOpen
	\bibfield  {author} {\bibinfo {author} {\bibfnamefont {P.}~\bibnamefont
			{Cieśliński}}, \bibinfo {author} {\bibfnamefont {W.}~\bibnamefont
			{Kłobus}}, \bibinfo {author} {\bibfnamefont {P.}~\bibnamefont {Kurzyński}},
		\bibinfo {author} {\bibfnamefont {T.}~\bibnamefont {Paterek}},\ and\ \bibinfo
		{author} {\bibfnamefont {W.}~\bibnamefont {Laskowski}},\ }\bibfield  {title}
	{\bibinfo {title} {The fastest generation of multipartite entanglement with
			natural interactions},\ }\href {https://doi.org/10.1088/1367-2630/acf953}
	{\bibfield  {journal} {\bibinfo  {journal} {New J. Phys.}\ }\textbf {\bibinfo
			{volume} {25}},\ \bibinfo {pages} {093040} (\bibinfo {year}
		{2023})}\BibitemShut {NoStop}%
	\bibitem [{\citenamefont {Facchi}\ \emph {et~al.}(2011)\citenamefont {Facchi},
		\citenamefont {Florio}, \citenamefont {Pascazio},\ and\ \citenamefont
		{Pepe}}]{PhysRevLett.107.260502}%
	\BibitemOpen
	\bibfield  {author} {\bibinfo {author} {\bibfnamefont {P.}~\bibnamefont
			{Facchi}}, \bibinfo {author} {\bibfnamefont {G.}~\bibnamefont {Florio}},
		\bibinfo {author} {\bibfnamefont {S.}~\bibnamefont {Pascazio}},\ and\
		\bibinfo {author} {\bibfnamefont {F.~V.}\ \bibnamefont {Pepe}},\ }\bibfield
	{title} {\bibinfo {title} {{G}reenberger-{H}orne-{Z}eilinger states and
			few-body {H}amiltonians},\ }\href
	{https://doi.org/10.1103/PhysRevLett.107.260502} {\bibfield  {journal}
		{\bibinfo  {journal} {Phys. Rev. Lett.}\ }\textbf {\bibinfo {volume} {107}},\
		\bibinfo {pages} {260502} (\bibinfo {year} {2011})}\BibitemShut {NoStop}%
	\bibitem [{\citenamefont {Dai}\ \emph {et~al.}(2017)\citenamefont {Dai},
		\citenamefont {Yang}, \citenamefont {Reingruber}, \citenamefont {Sun},
		\citenamefont {Xu}, \citenamefont {Chen}, \citenamefont {Yuan},\ and\
		\citenamefont {Pan}}]{Dai2017FourBody}%
	\BibitemOpen
	\bibfield  {author} {\bibinfo {author} {\bibfnamefont {H.-N.}\ \bibnamefont
			{Dai}}, \bibinfo {author} {\bibfnamefont {B.}~\bibnamefont {Yang}}, \bibinfo
		{author} {\bibfnamefont {A.}~\bibnamefont {Reingruber}}, \bibinfo {author}
		{\bibfnamefont {H.}~\bibnamefont {Sun}}, \bibinfo {author} {\bibfnamefont
			{X.-F.}\ \bibnamefont {Xu}}, \bibinfo {author} {\bibfnamefont {Y.-A.}\
			\bibnamefont {Chen}}, \bibinfo {author} {\bibfnamefont {Z.-S.}\ \bibnamefont
			{Yuan}},\ and\ \bibinfo {author} {\bibfnamefont {J.-W.}\ \bibnamefont
			{Pan}},\ }\bibfield  {title} {\bibinfo {title} {Four-body ring-exchange
			interactions and anyonic statistics within a minimal toric-code
			\uppercase{H}amiltonian},\ }\href {https://doi.org/10.1038/NPHYS4243}
	{\bibfield  {journal} {\bibinfo  {journal} {Nature Phys.}\ }\textbf {\bibinfo
			{volume} {13}},\ \bibinfo {pages} {1195–1200} (\bibinfo {year}
		{2017})}\BibitemShut {NoStop}%
	\bibitem [{\citenamefont {Menke}\ \emph {et~al.}(2022)\citenamefont {Menke},
		\citenamefont {Banner}, \citenamefont {Bergamaschi}, \citenamefont
		{Di~Paolo}, \citenamefont {Veps\"al\"ainen}, \citenamefont {Weber},
		\citenamefont {Winik}, \citenamefont {Melville}, \citenamefont {Niedzielski},
		\citenamefont {Rosenberg}, \citenamefont {Serniak}, \citenamefont {Schwartz},
		\citenamefont {Yoder}, \citenamefont {Aspuru-Guzik}, \citenamefont
		{Gustavsson}, \citenamefont {Grover}, \citenamefont {Hirjibehedin},
		\citenamefont {Kerman},\ and\ \citenamefont
		{Oliver}}]{Menke2022Demonstration}%
	\BibitemOpen
	\bibfield  {author} {\bibinfo {author} {\bibfnamefont {T.}~\bibnamefont
			{Menke}}, \bibinfo {author} {\bibfnamefont {W.~P.}\ \bibnamefont {Banner}},
		\bibinfo {author} {\bibfnamefont {T.~R.}\ \bibnamefont {Bergamaschi}},
		\bibinfo {author} {\bibfnamefont {A.}~\bibnamefont {Di~Paolo}}, \bibinfo
		{author} {\bibfnamefont {A.}~\bibnamefont {Veps\"al\"ainen}}, \bibinfo
		{author} {\bibfnamefont {S.~J.}\ \bibnamefont {Weber}}, \bibinfo {author}
		{\bibfnamefont {R.}~\bibnamefont {Winik}}, \bibinfo {author} {\bibfnamefont
			{A.}~\bibnamefont {Melville}}, \bibinfo {author} {\bibfnamefont {B.~M.}\
			\bibnamefont {Niedzielski}}, \bibinfo {author} {\bibfnamefont
			{D.}~\bibnamefont {Rosenberg}}, \bibinfo {author} {\bibfnamefont
			{K.}~\bibnamefont {Serniak}}, \bibinfo {author} {\bibfnamefont {M.~E.}\
			\bibnamefont {Schwartz}}, \bibinfo {author} {\bibfnamefont {J.~L.}\
			\bibnamefont {Yoder}}, \bibinfo {author} {\bibfnamefont {A.}~\bibnamefont
			{Aspuru-Guzik}}, \bibinfo {author} {\bibfnamefont {S.}~\bibnamefont
			{Gustavsson}}, \bibinfo {author} {\bibfnamefont {J.~A.}\ \bibnamefont
			{Grover}}, \bibinfo {author} {\bibfnamefont {C.~F.}\ \bibnamefont
			{Hirjibehedin}}, \bibinfo {author} {\bibfnamefont {A.~J.}\ \bibnamefont
			{Kerman}},\ and\ \bibinfo {author} {\bibfnamefont {W.~D.}\ \bibnamefont
			{Oliver}},\ }\bibfield  {title} {\bibinfo {title} {Demonstration of tunable
			three-body interactions between superconducting qubits},\ }\href
	{https://doi.org/10.1103/PhysRevLett.129.220501} {\bibfield  {journal}
		{\bibinfo  {journal} {Phys. Rev. Lett.}\ }\textbf {\bibinfo {volume} {129}},\
		\bibinfo {pages} {220501} (\bibinfo {year} {2022})}\BibitemShut {NoStop}%
	\bibitem [{\citenamefont {Katz}\ \emph {et~al.}(2023)\citenamefont {Katz},
		\citenamefont {Feng}, \citenamefont {Risinger}, \citenamefont {Monroe},\ and\
		\citenamefont {Cetina}}]{Katz2023Demonstration}%
	\BibitemOpen
	\bibfield  {author} {\bibinfo {author} {\bibfnamefont {O.}~\bibnamefont
			{Katz}}, \bibinfo {author} {\bibfnamefont {L.}~\bibnamefont {Feng}}, \bibinfo
		{author} {\bibfnamefont {A.}~\bibnamefont {Risinger}}, \bibinfo {author}
		{\bibfnamefont {C.}~\bibnamefont {Monroe}},\ and\ \bibinfo {author}
		{\bibfnamefont {M.}~\bibnamefont {Cetina}},\ }\bibfield  {title} {\bibinfo
		{title} {Demonstration of three- and four-body interactions between
			trapped-ion spins},\ }\href {https://doi.org/10.1038/s41567-023-02102-7}
	{\bibfield  {journal} {\bibinfo  {journal} {Nature Phys.}\ }\textbf {\bibinfo
			{volume} {19}},\ \bibinfo {pages} {1452–1458} (\bibinfo {year}
		{2023})}\BibitemShut {NoStop}%
	\bibitem [{\citenamefont {Nill}\ \emph {et~al.}(2025)\citenamefont {Nill},
		\citenamefont {de~L\'es\'eleuc}, \citenamefont {Gro\ss{}},\ and\
		\citenamefont {Lesanovsky}}]{Nilss2025Resonant}%
	\BibitemOpen
	\bibfield  {author} {\bibinfo {author} {\bibfnamefont {C.}~\bibnamefont
			{Nill}}, \bibinfo {author} {\bibfnamefont {S.}~\bibnamefont
			{de~L\'es\'eleuc}}, \bibinfo {author} {\bibfnamefont {C.}~\bibnamefont
			{Gro\ss{}}},\ and\ \bibinfo {author} {\bibfnamefont {I.}~\bibnamefont
			{Lesanovsky}},\ }\bibfield  {title} {\bibinfo {title} {Resonant stroboscopic
			\uppercase{R}ydberg dressing: Electron-motion coupling and multibody
			interactions},\ }\href {https://doi.org/10.1103/PhysRevA.111.L041104}
	{\bibfield  {journal} {\bibinfo  {journal} {Phys. Rev. A}\ }\textbf {\bibinfo
			{volume} {111}},\ \bibinfo {pages} {L041104} (\bibinfo {year}
		{2025})}\BibitemShut {NoStop}%
	\bibitem [{\citenamefont {Wang}\ \emph {et~al.}(2001)\citenamefont {Wang},
		\citenamefont {S{\o}rensen},\ and\ \citenamefont {M{\o}lmer}}]{Wang2001}%
	\BibitemOpen
	\bibfield  {author} {\bibinfo {author} {\bibfnamefont {X.}~\bibnamefont
			{Wang}}, \bibinfo {author} {\bibfnamefont {A.}~\bibnamefont {S{\o}rensen}},\
		and\ \bibinfo {author} {\bibfnamefont {K.}~\bibnamefont {M{\o}lmer}},\
	}\bibfield  {title} {\bibinfo {title} {Multibit gates for quantum
			computing},\ }\href
	{https://journals.aps.org/prl/abstract/10.1103/PhysRevLett.86.3907}
	{\bibfield  {journal} {\bibinfo  {journal} {Phys. Rev. Lett.}\ }\textbf
		{\bibinfo {volume} {86}},\ \bibinfo {pages} {3907} (\bibinfo {year}
		{2001})}\BibitemShut {NoStop}%
	\bibitem [{\citenamefont {Ezawa}(2024)}]{Ezawa2024Systematic}%
	\BibitemOpen
	\bibfield  {author} {\bibinfo {author} {\bibfnamefont {M.}~\bibnamefont
			{Ezawa}},\ }\bibfield  {title} {\bibinfo {title} {Systematic construction of
			topological-nontopological hybrid universal quantum gates based on many-body
			\uppercase{M}ajorana fermion interactions},\ }\href
	{https://doi.org/10.1103/PhysRevB.110.045417} {\bibfield  {journal} {\bibinfo
			{journal} {Phys. Rev. B}\ }\textbf {\bibinfo {volume} {110}},\ \bibinfo
		{pages} {045417} (\bibinfo {year} {2024})}\BibitemShut {NoStop}%
	\bibitem [{\citenamefont {Terhal}(2015)}]{RevModPhys.87.307}%
	\BibitemOpen
	\bibfield  {author} {\bibinfo {author} {\bibfnamefont {B.~M.}\ \bibnamefont
			{Terhal}},\ }\bibfield  {title} {\bibinfo {title} {Quantum error correction
			for quantum memories},\ }\href {https://doi.org/10.1103/RevModPhys.87.307}
	{\bibfield  {journal} {\bibinfo  {journal} {Rev. Mod. Phys.}\ }\textbf
		{\bibinfo {volume} {87}},\ \bibinfo {pages} {307} (\bibinfo {year}
		{2015})}\BibitemShut {NoStop}%
	\bibitem [{\citenamefont {Kitaev}(2003)}]{Kitaev2003}%
	\BibitemOpen
	\bibfield  {author} {\bibinfo {author} {\bibfnamefont {A.~Y.}\ \bibnamefont
			{Kitaev}},\ }\bibfield  {title} {\bibinfo {title} {Fault-tolerant quantum
			computation by anyons},\ }\href
	{https://www.sciencedirect.com/science/article/pii/S0003491602000180?casa_token=7nKrad-p28MAAAAA:9PHUSdk8bc8LezRIRcKUNFJBkhTuEbWLIOuddMgh_cPAWni4uWedZAKGuPJg9JcG0VNSOO12O6pA}
	{\bibfield  {journal} {\bibinfo  {journal} {Ann. Phys.}\ }\textbf {\bibinfo
			{volume} {303}},\ \bibinfo {pages} {2–30} (\bibinfo {year}
		{2003})}\BibitemShut {NoStop}%
	\bibitem [{\citenamefont {Vy}\ \emph {et~al.}(2013)\citenamefont {Vy},
		\citenamefont {Wang},\ and\ \citenamefont {Jacobs}}]{Vy2013ErrorTransparent}%
	\BibitemOpen
	\bibfield  {author} {\bibinfo {author} {\bibfnamefont {O.}~\bibnamefont
			{Vy}}, \bibinfo {author} {\bibfnamefont {X.}~\bibnamefont {Wang}},\ and\
		\bibinfo {author} {\bibfnamefont {K.}~\bibnamefont {Jacobs}},\ }\bibfield
	{title} {\bibinfo {title} {Error-transparent evolution: the ability of
			multi-body interactions to bypass dechorence},\ }\href
	{https://doi.org/10.1088/1367-2630/15/5/053002} {\bibfield  {journal}
		{\bibinfo  {journal} {New. J. Phys.}\ }\textbf {\bibinfo {volume} {15}},\
		\bibinfo {pages} {053002} (\bibinfo {year} {2013})}\BibitemShut {NoStop}%
	\bibitem [{\citenamefont {Kawaguchi}\ and\ \citenamefont
		{Ueda}(2012)}]{Kawaguchi_2012}%
	\BibitemOpen
	\bibfield  {author} {\bibinfo {author} {\bibfnamefont {Y.}~\bibnamefont
			{Kawaguchi}}\ and\ \bibinfo {author} {\bibfnamefont {M.}~\bibnamefont
			{Ueda}},\ }\bibfield  {title} {\bibinfo {title} {Spinor {B}ose–{E}instein
			condensates},\ }\href {https://doi.org/10.1016/j.physrep.2012.07.005}
	{\bibfield  {journal} {\bibinfo  {journal} {Phys. Rep.}\ }\textbf {\bibinfo
			{volume} {520}},\ \bibinfo {pages} {253–381} (\bibinfo {year}
		{2012})}\BibitemShut {NoStop}%
	\bibitem [{\citenamefont {Stamper-Kurn}\ and\ \citenamefont
		{Ueda}(2013)}]{Stamper2013}%
	\BibitemOpen
	\bibfield  {author} {\bibinfo {author} {\bibfnamefont {D.~M.}\ \bibnamefont
			{Stamper-Kurn}}\ and\ \bibinfo {author} {\bibfnamefont {M.}~\bibnamefont
			{Ueda}},\ }\bibfield  {title} {\bibinfo {title} {Spinor {B}ose gases:
			Symmetries, magnetism, and quantum dynamics},\ }\href
	{https://doi.org/10.1103/RevModPhys.85.1191} {\bibfield  {journal} {\bibinfo
			{journal} {Rev. Mod. Phys.}\ }\textbf {\bibinfo {volume} {85}},\ \bibinfo
		{pages} {1191} (\bibinfo {year} {2013})}\BibitemShut {NoStop}%
	\bibitem [{\citenamefont {Chen}\ \emph {et~al.}(2019)\citenamefont {Chen},
		\citenamefont {Tang}, \citenamefont {Austin}, \citenamefont {Shaw},
		\citenamefont {Zhao},\ and\ \citenamefont {Liu}}]{Zihe2019}%
	\BibitemOpen
	\bibfield  {author} {\bibinfo {author} {\bibfnamefont {Z.}~\bibnamefont
			{Chen}}, \bibinfo {author} {\bibfnamefont {T.}~\bibnamefont {Tang}}, \bibinfo
		{author} {\bibfnamefont {J.}~\bibnamefont {Austin}}, \bibinfo {author}
		{\bibfnamefont {Z.}~\bibnamefont {Shaw}}, \bibinfo {author} {\bibfnamefont
			{L.}~\bibnamefont {Zhao}},\ and\ \bibinfo {author} {\bibfnamefont
			{Y.}~\bibnamefont {Liu}},\ }\bibfield  {title} {\bibinfo {title} {Quantum
			quench and nonequilibrium dynamics in lattice-confined spinor condensates},\
	}\href {https://doi.org/10.1103/PhysRevLett.123.113002} {\bibfield  {journal}
		{\bibinfo  {journal} {Phys. Rev. Lett.}\ }\textbf {\bibinfo {volume} {123}},\
		\bibinfo {pages} {113002} (\bibinfo {year} {2019})},\ \bibinfo {note} {and
		references therein}\BibitemShut {NoStop}%
	\bibitem [{\citenamefont {Austin}\ \emph
		{et~al.}(2021{\natexlab{a}})\citenamefont {Austin}, \citenamefont {Shaw},
		\citenamefont {Chen}, \citenamefont {Mahmud},\ and\ \citenamefont
		{Liu}}]{Jared2021Manipulation}%
	\BibitemOpen
	\bibfield  {author} {\bibinfo {author} {\bibfnamefont {J.~O.}\ \bibnamefont
			{Austin}}, \bibinfo {author} {\bibfnamefont {Z.~N.}\ \bibnamefont {Shaw}},
		\bibinfo {author} {\bibfnamefont {Z.}~\bibnamefont {Chen}}, \bibinfo {author}
		{\bibfnamefont {K.~W.}\ \bibnamefont {Mahmud}},\ and\ \bibinfo {author}
		{\bibfnamefont {Y.}~\bibnamefont {Liu}},\ }\bibfield  {title} {\bibinfo
		{title} {Manipulating atom-number distributions and detecting spatial
			distributions in lattice-confined spinor gases},\ }\href
	{https://doi.org/10.1103/PhysRevA.104.L041304} {\bibfield  {journal}
		{\bibinfo  {journal} {Phys. Rev. A}\ }\textbf {\bibinfo {volume} {104}},\
		\bibinfo {pages} {L041304} (\bibinfo {year} {2021}{\natexlab{a}})},\ \bibinfo
	{note} {and references therein}\BibitemShut {NoStop}%
	\bibitem [{\citenamefont {Austin}\ \emph
		{et~al.}(2021{\natexlab{b}})\citenamefont {Austin}, \citenamefont {Chen},
		\citenamefont {Shaw}, \citenamefont {Mahmud},\ and\ \citenamefont
		{Liu}}]{Jared2021Quantum}%
	\BibitemOpen
	\bibfield  {author} {\bibinfo {author} {\bibfnamefont {J.~O.}~\bibnamefont
			{Austin}}, \bibinfo {author} {\bibfnamefont {Z.}~\bibnamefont {Chen}},
		\bibinfo {author} {\bibfnamefont {Z.~N.}\ \bibnamefont {Shaw}}, \bibinfo
		{author} {\bibfnamefont {K.~W.}\ \bibnamefont {Mahmud}},\ and\ \bibinfo
		{author} {\bibfnamefont {Y.}~\bibnamefont {Liu}},\ }\bibfield  {title}
	{\bibinfo {title} {Quantum critical dynamics in a spinor {H}ubbard model
			quantum simulator},\ }\href
	{https://doi.org/https://doi.org/10.1038/s42005-021-00562-y} {\bibfield
		{journal} {\bibinfo  {journal} {Comm. Phys.}\ }\textbf {\bibinfo {volume}
			{4}},\ \bibinfo {pages} {61} (\bibinfo {year}
		{2021}{\natexlab{b}})}\BibitemShut {NoStop}%
	\bibitem [{\citenamefont {Hardesty-Shaw}\ \emph
		{et~al.}(2023{\natexlab{a}})\citenamefont {Hardesty-Shaw}, \citenamefont
		{Guan}, \citenamefont {Austin}, \citenamefont {Blume}, \citenamefont
		{Lewis-Swan},\ and\ \citenamefont {Liu}}]{Zach1}%
	\BibitemOpen
	\bibfield  {author} {\bibinfo {author} {\bibfnamefont {Z.~N.}\ \bibnamefont
			{Hardesty-Shaw}}, \bibinfo {author} {\bibfnamefont {Q.}~\bibnamefont {Guan}},
		\bibinfo {author} {\bibfnamefont {J.~O.}\ \bibnamefont {Austin}}, \bibinfo
		{author} {\bibfnamefont {D.}~\bibnamefont {Blume}}, \bibinfo {author}
		{\bibfnamefont {R.~J.}\ \bibnamefont {Lewis-Swan}},\ and\ \bibinfo {author}
		{\bibfnamefont {Y.}~\bibnamefont {Liu}},\ }\bibfield  {title} {\bibinfo
		{title} {Quench-induced nonequilibrium dynamics of spinor gases in a moving
			lattice},\ }\href {https://doi.org/10.1103/PhysRevA.107.053311} {\bibfield
		{journal} {\bibinfo  {journal} {Phys. Rev. A}\ }\textbf {\bibinfo {volume}
			{107}},\ \bibinfo {pages} {053311} (\bibinfo {year}
		{2023}{\natexlab{a}})}\BibitemShut {NoStop}%
	\bibitem [{\citenamefont {Austin-Harris}\ \emph {et~al.}(2024)\citenamefont
		{Austin-Harris}, \citenamefont {Hardesty-Shaw}, \citenamefont {Guan},
		\citenamefont {Binegar}, \citenamefont {Blume}, \citenamefont {Lewis-Swan},\
		and\ \citenamefont {Liu}}]{Jared2024}%
	\BibitemOpen
	\bibfield  {author} {\bibinfo {author} {\bibfnamefont {J.~O.}\ \bibnamefont
			{Austin-Harris}}, \bibinfo {author} {\bibfnamefont {Z.~N.}\ \bibnamefont
			{Hardesty-Shaw}}, \bibinfo {author} {\bibfnamefont {Q.}~\bibnamefont {Guan}},
		\bibinfo {author} {\bibfnamefont {C.}~\bibnamefont {Binegar}}, \bibinfo
		{author} {\bibfnamefont {D.}~\bibnamefont {Blume}}, \bibinfo {author}
		{\bibfnamefont {R.~J.}\ \bibnamefont {Lewis-Swan}},\ and\ \bibinfo {author}
		{\bibfnamefont {Y.}~\bibnamefont {Liu}},\ }\bibfield  {title} {\bibinfo
		{title} {Engineering dynamical phase diagrams with driven lattices in spinor
			gases},\ }\href {https://doi.org/10.1103/PhysRevA.109.043309} {\bibfield
		{journal} {\bibinfo  {journal} {Phys. Rev. A}\ }\textbf {\bibinfo {volume}
			{109}},\ \bibinfo {pages} {043309} (\bibinfo {year} {2024})}\BibitemShut
	{NoStop}%
	\bibitem [{\citenamefont {Jiang}\ \emph {et~al.}(2016)\citenamefont {Jiang},
		\citenamefont {Zhao}, \citenamefont {Wang}, \citenamefont {Chen},
		\citenamefont {Tang}, \citenamefont {Duan},\ and\ \citenamefont
		{Liu}}]{Jie2016}%
	\BibitemOpen
	\bibfield  {author} {\bibinfo {author} {\bibfnamefont {J.}~\bibnamefont
			{Jiang}}, \bibinfo {author} {\bibfnamefont {L.}~\bibnamefont {Zhao}},
		\bibinfo {author} {\bibfnamefont {S.-T.}\ \bibnamefont {Wang}}, \bibinfo
		{author} {\bibfnamefont {Z.}~\bibnamefont {Chen}}, \bibinfo {author}
		{\bibfnamefont {T.}~\bibnamefont {Tang}}, \bibinfo {author} {\bibfnamefont
			{L.-M.}\ \bibnamefont {Duan}},\ and\ \bibinfo {author} {\bibfnamefont
			{Y.}~\bibnamefont {Liu}},\ }\bibfield  {title} {\bibinfo {title} {First-order
			superfluid-to-{M}ott-insulator phase transitions in spinor condensates},\
	}\href {https://doi.org/10.1103/PhysRevA.93.063607} {\bibfield  {journal}
		{\bibinfo  {journal} {Phys. Rev. A}\ }\textbf {\bibinfo {volume} {93}},\
		\bibinfo {pages} {063607} (\bibinfo {year} {2016})}\BibitemShut {NoStop}%
	\bibitem [{\citenamefont {Hardesty-Shaw}\ \emph
		{et~al.}(2023{\natexlab{b}})\citenamefont {Hardesty-Shaw}, \citenamefont
		{Guan}, \citenamefont {Austin}, \citenamefont {Blume}, \citenamefont
		{Lewis-Swan},\ and\ \citenamefont {Liu}}]{Zach2}%
	\BibitemOpen
	\bibfield  {author} {\bibinfo {author} {\bibfnamefont {Z.~N.}\ \bibnamefont
			{Hardesty-Shaw}}, \bibinfo {author} {\bibfnamefont {Q.}~\bibnamefont {Guan}},
		\bibinfo {author} {\bibfnamefont {J.~O.}\ \bibnamefont {Austin}}, \bibinfo
		{author} {\bibfnamefont {D.}~\bibnamefont {Blume}}, \bibinfo {author}
		{\bibfnamefont {R.~J.}\ \bibnamefont {Lewis-Swan}},\ and\ \bibinfo {author}
		{\bibfnamefont {Y.}~\bibnamefont {Liu}},\ }\bibfield  {title} {\bibinfo
		{title} {Nonlinear multistate tunneling dynamics in a spinor
			{B}ose-{E}instein condensate},\ }\href
	{https://doi.org/10.1103/PhysRevA.108.053307} {\bibfield  {journal} {\bibinfo
			{journal} {Phys. Rev. A}\ }\textbf {\bibinfo {volume} {108}},\ \bibinfo
		{pages} {053307} (\bibinfo {year} {2023}{\natexlab{b}})}\BibitemShut
	{NoStop}%
	\bibitem [{\citenamefont {Nabi}(2023)}]{Nabi2022Interplay}%
	\BibitemOpen
	\bibfield  {author} {\bibinfo {author} {\bibfnamefont {S.~N.}\ \bibnamefont
			{Nabi}},\ }\bibfield  {title} {\bibinfo {title} {Interplay of the staggered
			and three-body interaction potentials on the quantum phases of a spin-1
			ultracold atom in an optical lattice},\ }\href
	{https://doi.org/https://doi.org/10.1002/andp.202200482} {\bibfield
		{journal} {\bibinfo  {journal} {Annalen der Physik}\ }\textbf {\bibinfo
			{volume} {535}},\ \bibinfo {pages} {2200482} (\bibinfo {year}
		{2023})}\BibitemShut {NoStop}%
	\bibitem [{\citenamefont {Nabi}\ and\ \citenamefont
		{Basu}(2018)}]{nabi2018quantum}%
	\BibitemOpen
	\bibfield  {author} {\bibinfo {author} {\bibfnamefont {S.~N.}\ \bibnamefont
			{Nabi}}\ and\ \bibinfo {author} {\bibfnamefont {S.}~\bibnamefont {Basu}},\
	}\bibfield  {title} {\bibinfo {title} {Quantum phases of a spin-1 ultracold
			{B}ose gas with three-body interactions},\ }\href
	{https://doi.org/10.1209/0295-5075/121/46002} {\bibfield  {journal} {\bibinfo
			{journal} {Europhys. Lett.}\ }\textbf {\bibinfo {volume} {121}},\ \bibinfo
		{pages} {46002} (\bibinfo {year} {2018})}\BibitemShut {NoStop}%
	\bibitem [{\citenamefont {Hincapie-F}\ \emph {et~al.}(2018)\citenamefont
		{Hincapie-F}, \citenamefont {Franco},\ and\ \citenamefont
		{Silva-Valencia}}]{hincapie2018spin}%
	\BibitemOpen
	\bibfield  {author} {\bibinfo {author} {\bibfnamefont {A.}~\bibnamefont
			{Hincapie-F}}, \bibinfo {author} {\bibfnamefont {R.}~\bibnamefont {Franco}},\
		and\ \bibinfo {author} {\bibfnamefont {J.}~\bibnamefont {Silva-Valencia}},\
	}\bibfield  {title} {\bibinfo {title} {Spin-1 {B}ose--{H}ubbard model with
			two- and three-body interactions},\ }\href
	{https://doi.org/10.1016/j.physleta.2018.04.029} {\bibfield  {journal}
		{\bibinfo  {journal} {Phys. Lett. A}\ }\textbf {\bibinfo {volume} {382}},\
		\bibinfo {pages} {1760–1765} (\bibinfo {year} {2018})}\BibitemShut
	{NoStop}%
	\bibitem [{\citenamefont {Mestrom}\ \emph {et~al.}(2021)\citenamefont
		{Mestrom}, \citenamefont {Li}, \citenamefont {Colussi}, \citenamefont
		{Secker},\ and\ \citenamefont {Kokkelmans}}]{Mestrom2021Mixing}%
	\BibitemOpen
	\bibfield  {author} {\bibinfo {author} {\bibfnamefont {P.~M.~A.}\
			\bibnamefont {Mestrom}}, \bibinfo {author} {\bibfnamefont {J.-L.}\
			\bibnamefont {Li}}, \bibinfo {author} {\bibfnamefont {V.~E.}\ \bibnamefont
			{Colussi}}, \bibinfo {author} {\bibfnamefont {T.}~\bibnamefont {Secker}},\
		and\ \bibinfo {author} {\bibfnamefont {S.~J. J. M.~F.}\ \bibnamefont
			{Kokkelmans}},\ }\bibfield  {title} {\bibinfo {title} {Three-body spin mixing
			in spin-1 {B}ose-{E}instein condensates},\ }\href
	{https://doi.org/10.1103/PhysRevA.104.023321} {\bibfield  {journal} {\bibinfo
			{journal} {Phys. Rev. A}\ }\textbf {\bibinfo {volume} {104}},\ \bibinfo
		{pages} {023321} (\bibinfo {year} {2021})}\BibitemShut {NoStop}%
	\bibitem [{\citenamefont {Sun}\ \emph {et~al.}(2017)\citenamefont {Sun},
		\citenamefont {Xu}, \citenamefont {Pu},\ and\ \citenamefont
		{Zhang}}]{Sun2017Efficient}%
	\BibitemOpen
	\bibfield  {author} {\bibinfo {author} {\bibfnamefont {H.}~\bibnamefont
			{Sun}}, \bibinfo {author} {\bibfnamefont {P.}~\bibnamefont {Xu}}, \bibinfo
		{author} {\bibfnamefont {H.}~\bibnamefont {Pu}},\ and\ \bibinfo {author}
		{\bibfnamefont {W.}~\bibnamefont {Zhang}},\ }\bibfield  {title} {\bibinfo
		{title} {Efficient generation of many-body singlet states of spin-1 bosons in
			optical superlattices},\ }\href {https://doi.org/10.1103/PhysRevA.95.063624}
	{\bibfield  {journal} {\bibinfo  {journal} {Phys. Rev. A}\ }\textbf {\bibinfo
			{volume} {95}},\ \bibinfo {pages} {063624} (\bibinfo {year}
		{2017})}\BibitemShut {NoStop}%
	\bibitem [{\citenamefont {T\'oth}\ and\ \citenamefont
		{Mitchell}(2010)}]{Toth2010Generation}%
	\BibitemOpen
	\bibfield  {author} {\bibinfo {author} {\bibfnamefont {G.}~\bibnamefont
			{T\'oth}}\ and\ \bibinfo {author} {\bibfnamefont {M.~W.}\ \bibnamefont
			{Mitchell}},\ }\bibfield  {title} {\bibinfo {title} {Generation of
			macroscopic spin singlet states in atomic ensembles},\ }\href
	{https://doi.org/10.1088/1367-2630/12/5/053007} {\bibfield  {journal}
		{\bibinfo  {journal} {New J. Phys.}\ }\textbf {\bibinfo {volume} {12}},\
		\bibinfo {pages} {053007} (\bibinfo {year} {2010})}\BibitemShut {NoStop}%
	\bibitem [{\citenamefont {Urizar-Lanz}\ \emph {et~al.}(2013)\citenamefont
		{Urizar-Lanz}, \citenamefont {Hyllus}, \citenamefont {Egusquiza},
		\citenamefont {Mitchell},\ and\ \citenamefont
		{T\'oth}}]{urizar2013macroscopic}%
	\BibitemOpen
	\bibfield  {author} {\bibinfo {author} {\bibfnamefont {I.}~\bibnamefont
			{Urizar-Lanz}}, \bibinfo {author} {\bibfnamefont {P.}~\bibnamefont {Hyllus}},
		\bibinfo {author} {\bibfnamefont {I.~L.}\ \bibnamefont {Egusquiza}}, \bibinfo
		{author} {\bibfnamefont {M.~W.}\ \bibnamefont {Mitchell}},\ and\ \bibinfo
		{author} {\bibfnamefont {G.}~\bibnamefont {T\'oth}},\ }\bibfield  {title}
	{\bibinfo {title} {Macroscopic singlet states for gradient magnetometry},\
	}\href {https://doi.org/10.1103/PhysRevA.88.013626} {\bibfield  {journal}
		{\bibinfo  {journal} {Phys. Rev. A}\ }\textbf {\bibinfo {volume} {88}},\
		\bibinfo {pages} {013626} (\bibinfo {year} {2013})}\BibitemShut {NoStop}%
	\bibitem [{\citenamefont {Behbood}\ \emph {et~al.}(2014)\citenamefont
		{Behbood}, \citenamefont {Martin~Ciurana}, \citenamefont {Colangelo},
		\citenamefont {Napolitano}, \citenamefont {T\'oth}, \citenamefont {Sewell},\
		and\ \citenamefont {Mitchell}}]{behbood2014generation}%
	\BibitemOpen
	\bibfield  {author} {\bibinfo {author} {\bibfnamefont {N.}~\bibnamefont
			{Behbood}}, \bibinfo {author} {\bibfnamefont {F.}~\bibnamefont
			{Martin~Ciurana}}, \bibinfo {author} {\bibfnamefont {G.}~\bibnamefont
			{Colangelo}}, \bibinfo {author} {\bibfnamefont {M.}~\bibnamefont
			{Napolitano}}, \bibinfo {author} {\bibfnamefont {G.}~\bibnamefont {T\'oth}},
		\bibinfo {author} {\bibfnamefont {R.~J.}\ \bibnamefont {Sewell}},\ and\
		\bibinfo {author} {\bibfnamefont {M.~W.}\ \bibnamefont {Mitchell}},\
	}\bibfield  {title} {\bibinfo {title} {Generation of macroscopic singlet
			states in a cold atomic ensemble},\ }\href
	{https://doi.org/10.1103/PhysRevLett.113.093601} {\bibfield  {journal}
		{\bibinfo  {journal} {Phys. Rev. Lett.}\ }\textbf {\bibinfo {volume} {113}},\
		\bibinfo {pages} {093601} (\bibinfo {year} {2014})}\BibitemShut {NoStop}%
	\bibitem [{\citenamefont {Evrard}\ \emph {et~al.}(2021)\citenamefont {Evrard},
		\citenamefont {Qu}, \citenamefont {Dalibard},\ and\ \citenamefont
		{Gerbier}}]{Evrard2021Observation}%
	\BibitemOpen
	\bibfield  {author} {\bibinfo {author} {\bibfnamefont {B.}~\bibnamefont
			{Evrard}}, \bibinfo {author} {\bibfnamefont {A.}~\bibnamefont {Qu}}, \bibinfo
		{author} {\bibfnamefont {J.}~\bibnamefont {Dalibard}},\ and\ \bibinfo
		{author} {\bibfnamefont {F.}~\bibnamefont {Gerbier}},\ }\bibfield  {title}
	{\bibinfo {title} {Observation of fragmentation of a spinor {B}ose-{E}instein
			condensate},\ }\href {https://doi.org/10.1126/science.abd8206} {\bibfield
		{journal} {\bibinfo  {journal} {Science}\ }\textbf {\bibinfo {volume}
			{373}},\ \bibinfo {pages} {1340} (\bibinfo {year} {2021})}\BibitemShut
	{NoStop}%
	\bibitem [{SM()}]{SM}%
	\BibitemOpen
	\href@noop {} {\bibinfo {title} {{See Supplemental Material for additional
				details of our experimental procedures and theoretical models, which includes
				Ref. \cite{Mogensen2018}.}}}\BibitemShut {Stop}%
	\bibitem [{\citenamefont {Zhao}\ \emph {et~al.}(2014)\citenamefont {Zhao},
		\citenamefont {Jiang}, \citenamefont {Tang}, \citenamefont {Webb},\ and\
		\citenamefont {Liu}}]{Lichao2014}%
	\BibitemOpen
	\bibfield  {author} {\bibinfo {author} {\bibfnamefont {L.}~\bibnamefont
			{Zhao}}, \bibinfo {author} {\bibfnamefont {J.}~\bibnamefont {Jiang}},
		\bibinfo {author} {\bibfnamefont {T.}~\bibnamefont {Tang}}, \bibinfo {author}
		{\bibfnamefont {M.}~\bibnamefont {Webb}},\ and\ \bibinfo {author}
		{\bibfnamefont {Y.}~\bibnamefont {Liu}},\ }\bibfield  {title} {\bibinfo
		{title} {Dynamics in spinor condensates tuned by a microwave dressing
			field},\ }\href {https://doi.org/10.1103/PhysRevA.89.023608} {\bibfield
		{journal} {\bibinfo  {journal} {Phys. Rev. A}\ }\textbf {\bibinfo {volume}
			{89}},\ \bibinfo {pages} {023608} (\bibinfo {year} {2014})}\BibitemShut
	{NoStop}%
	\bibitem [{\citenamefont {Zhao}\ \emph {et~al.}(2015)\citenamefont {Zhao},
		\citenamefont {Jiang}, \citenamefont {Tang}, \citenamefont {Webb},\ and\
		\citenamefont {Liu}}]{Lichao2015}%
	\BibitemOpen
	\bibfield  {author} {\bibinfo {author} {\bibfnamefont {L.}~\bibnamefont
			{Zhao}}, \bibinfo {author} {\bibfnamefont {J.}~\bibnamefont {Jiang}},
		\bibinfo {author} {\bibfnamefont {T.}~\bibnamefont {Tang}}, \bibinfo {author}
		{\bibfnamefont {M.}~\bibnamefont {Webb}},\ and\ \bibinfo {author}
		{\bibfnamefont {Y.}~\bibnamefont {Liu}},\ }\bibfield  {title} {\bibinfo
		{title} {Antiferromagnetic spinor condensates in a two-dimensional optical
			lattice},\ }\href {https://doi.org/10.1103/PhysRevLett.114.225302} {\bibfield
		{journal} {\bibinfo  {journal} {Phys. Rev. Lett.}\ }\textbf {\bibinfo
			{volume} {114}},\ \bibinfo {pages} {225302} (\bibinfo {year}
		{2015})}\BibitemShut {NoStop}%
	\bibitem [{\citenamefont {Jiang}\ \emph {et~al.}(2014)\citenamefont {Jiang},
		\citenamefont {Zhao}, \citenamefont {Webb},\ and\ \citenamefont
		{Liu}}]{Jie2014}%
	\BibitemOpen
	\bibfield  {author} {\bibinfo {author} {\bibfnamefont {J.}~\bibnamefont
			{Jiang}}, \bibinfo {author} {\bibfnamefont {L.}~\bibnamefont {Zhao}},
		\bibinfo {author} {\bibfnamefont {M.}~\bibnamefont {Webb}},\ and\ \bibinfo
		{author} {\bibfnamefont {Y.}~\bibnamefont {Liu}},\ }\bibfield  {title}
	{\bibinfo {title} {Mapping the phase diagram of spinor condensates via
			adiabatic quantum phase transitions},\ }\href
	{https://doi.org/10.1103/PhysRevA.90.023610} {\bibfield  {journal} {\bibinfo
			{journal} {Phys. Rev. A}\ }\textbf {\bibinfo {volume} {90}},\ \bibinfo
		{pages} {023610} (\bibinfo {year} {2014})}\BibitemShut {NoStop}%
	\bibitem [{\citenamefont {Hammond}\ \emph {et~al.}(2022)\citenamefont
		{Hammond}, \citenamefont {Lavoine},\ and\ \citenamefont
		{Bourdel}}]{Hammond2022Tunable}%
	\BibitemOpen
	\bibfield  {author} {\bibinfo {author} {\bibfnamefont {A.}~\bibnamefont
			{Hammond}}, \bibinfo {author} {\bibfnamefont {L.}~\bibnamefont {Lavoine}},\
		and\ \bibinfo {author} {\bibfnamefont {T.}~\bibnamefont {Bourdel}},\
	}\bibfield  {title} {\bibinfo {title} {Tunable three-body interactions in
			driven two-component {B}ose-{E}instein condensates},\ }\href
	{https://doi.org/10.1103/PhysRevLett.128.083401} {\bibfield  {journal}
		{\bibinfo  {journal} {Phys. Rev. Lett.}\ }\textbf {\bibinfo {volume} {128}},\
		\bibinfo {pages} {083401} (\bibinfo {year} {2022})}\BibitemShut {NoStop}%
	\bibitem [{\citenamefont {Imambekov}\ \emph {et~al.}(2003)\citenamefont
		{Imambekov}, \citenamefont {Lukin},\ and\ \citenamefont
		{Demler}}]{Imambekov2003Spin}%
	\BibitemOpen
	\bibfield  {author} {\bibinfo {author} {\bibfnamefont {A.}~\bibnamefont
			{Imambekov}}, \bibinfo {author} {\bibfnamefont {M.}~\bibnamefont {Lukin}},\
		and\ \bibinfo {author} {\bibfnamefont {E.}~\bibnamefont {Demler}},\
	}\bibfield  {title} {\bibinfo {title} {Spin-exchange interactions of spin-one
			bosons in optical lattices: singlet, nematic, and dimerized phases},\ }\href
	{https://doi.org/10.1103/PhysRevA.68.063602} {\bibfield  {journal} {\bibinfo
			{journal} {Phys. Rev. A}\ }\textbf {\bibinfo {volume} {68}},\ \bibinfo
		{pages} {063602} (\bibinfo {year} {2003})}\BibitemShut {NoStop}%
	\bibitem [{\citenamefont {Pan}\ \emph {et~al.}(2019)\citenamefont {Pan},
		\citenamefont {Chen},\ and\ \citenamefont {Cui}}]{Pan2019HighOrder}%
	\BibitemOpen
	\bibfield  {author} {\bibinfo {author} {\bibfnamefont {L.}~\bibnamefont
			{Pan}}, \bibinfo {author} {\bibfnamefont {S.}~\bibnamefont {Chen}},\ and\
		\bibinfo {author} {\bibfnamefont {X.}~\bibnamefont {Cui}},\ }\bibfield
	{title} {\bibinfo {title} {High-order exceptional points in ultracold {B}ose
			gases},\ }\href {https://doi.org/10.1103/PhysRevA.99.011601} {\bibfield
		{journal} {\bibinfo  {journal} {Phys. Rev. A}\ }\textbf {\bibinfo {volume}
			{99}},\ \bibinfo {pages} {011601(R)} (\bibinfo {year} {2019})}\BibitemShut
	{NoStop}%
	\bibitem [{\citenamefont {Mogensen}\ and\ \citenamefont
		{Riseth}(2018)}]{Mogensen2018}%
	\BibitemOpen
	\bibfield  {author} {\bibinfo {author} {\bibfnamefont {P.~K.}\ \bibnamefont
			{Mogensen}}\ and\ \bibinfo {author} {\bibfnamefont {A.~N.}\ \bibnamefont
			{Riseth}},\ }\bibfield  {title} {\bibinfo {title} {Optim: A mathematical
			optimization package for {Julia}},\ }\href
	{https://doi.org/10.21105/joss.00615} {\bibfield  {journal} {\bibinfo
			{journal} {Journal of Open Source Software}\ }\textbf {\bibinfo {volume}
			{3}},\ \bibinfo {pages} {615} (\bibinfo {year} {2018})}\BibitemShut {NoStop}%
\end{thebibliography}
%

\end{document}


\title{Supplemental Materials for ``Lattice-enabled detection of spin-dependent three-body interactions"}
	\author{C. Binegar}
	\author{J. O. Austin-Harris}
    \affiliation{Department of Physics, Oklahoma State University, Stillwater, Oklahoma 74078, USA}
	\author{S. E. Begg}
    \affiliation{Department of Physics, Oklahoma State University, Stillwater, Oklahoma 74078, USA}
    \affiliation{Department of Physics, The University of Texas at Dallas, Richardson, Texas 75080, USA}
	\author{P. Sigdel}       
	\author{T. Bilitewski}
	\email{thomas.bilitewski@okstate.edu}
	\author{Y. Liu}
	\email{yingmei.liu@okstate.edu}
	\affiliation{Department of Physics, Oklahoma State University, Stillwater, Oklahoma 74078, USA}
	\maketitle
	\vspace{-2pc}
	\section{Experimental Details}
            Each experimental cycle begins by generating an $F=1$ spinor BEC of up to $1\times 10^5$ sodium atoms in a crossed optical dipole trap  with trapping frequencies $\omega_{x,y,z} \sim2\pi(130,130,180)~\mathrm{Hz}$. The atoms are prepared in their superfluid ground state, the longitudinally polarized (LP) state where $\rho_0 = 1$ and $M = 0$. Here $\rho_{m_F}$ is the fractional population of the $m_F$ hyperfine state and $M = \rho_1 - \rho_{-1}$ is the magnetization.  The atoms are then loaded into a cubic optical lattice with a lattice spacing of $\lambda/2$ and associated recoil energy $E_R=h^2 /(2M_{\mathrm{Na}}\lambda^2)$. Here $M_{\mathrm{Na}}$ is the atomic mass of sodium, $\lambda=1064~\mathrm{nm}$ is the wavelength of the lattice beams, and $h$ is the Planck constant. Similar to our prior work~\cite{Zihe2019}, nonequilibrium spin dynamics are initiated by two distinct quantum quench sequences. For one quench sequence, the lattice depth is quenched across the superfluid to Mott insulator phase transition to the final lattice depth $u_L$ while the quadratic Zeeman shift is held constant at $q$. In the other sequence, the lattice depth is adiabatically raised through the superfluid Mott insulator phase transition to $u_L$, before the quadratic Zeeman shift is quenched from a very high value to its final value $q$.  Both sequences result in few-body nonequilibrium dynamics which are monitored using a spin-resolved two-stage microwave time of flight imaging process~\cite{Zihe2019} after allowing the atoms to evolve in the lattice for a time $t_{\mathrm{hold}}$. 
 
	\section{Details on theoretical modeling}

\subsection{Determination of $U_2$, $V_2$ and $\chi_n$ from simulations}
\begin{figure} [h]
\begin{center}
\subfigure[$85$~Hz L-quench]
{\includegraphics[width=0.36\linewidth]{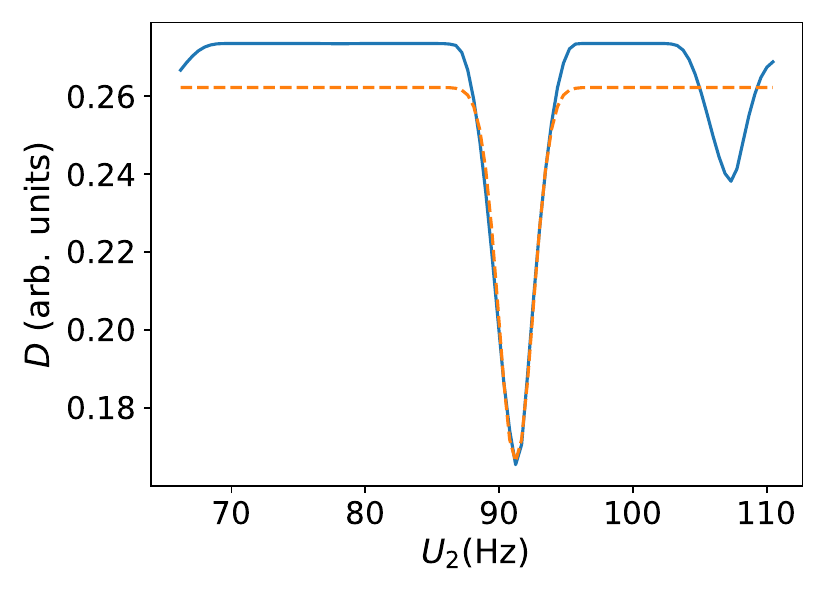}}
\subfigure[$44$~Hz Q-quench]{\includegraphics[width=0.36\linewidth]{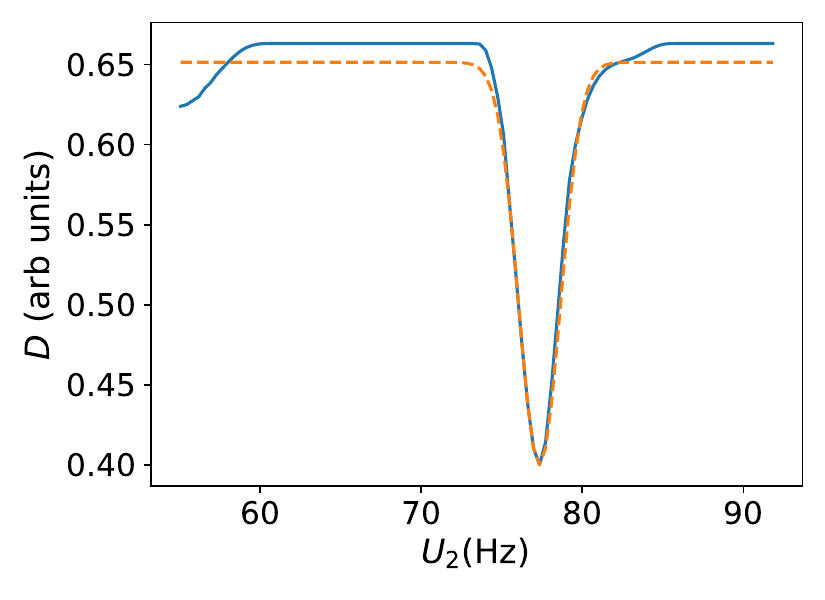}}
\subfigure[$85$~Hz L-quench]{\includegraphics[width=0.36\linewidth]{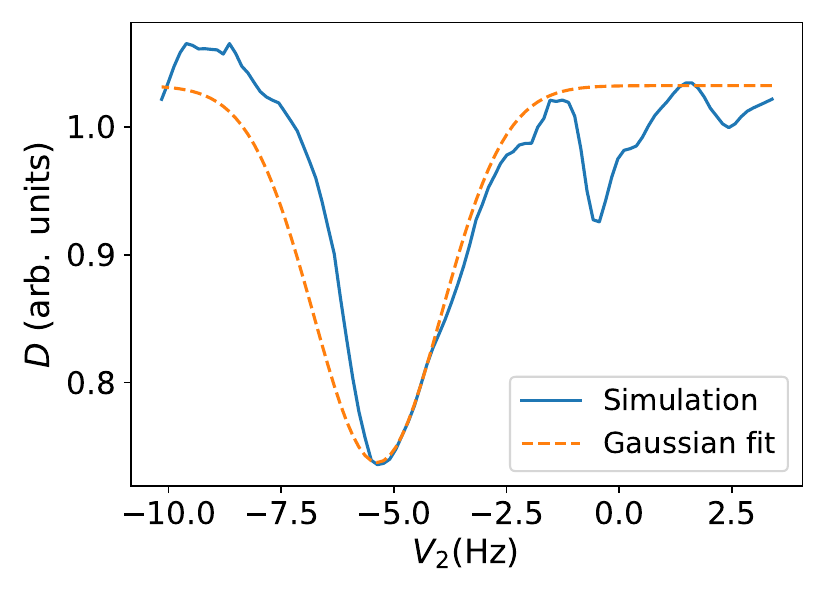}}
\subfigure[$44$~Hz Q-quench]{\includegraphics[width=0.36\linewidth]{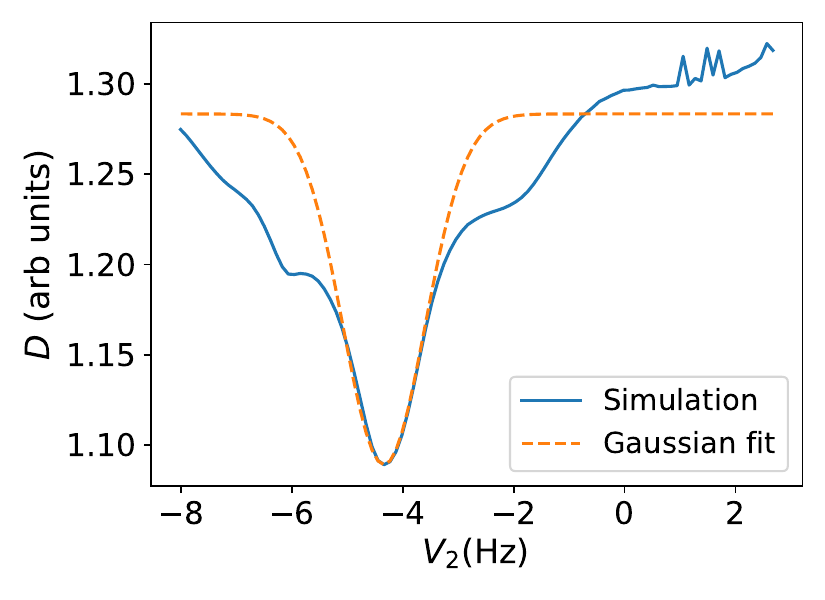}}
\end{center}
    \caption{Distance $D$ vs $U_2$ (a,c) and $V_2$ (b,d) for a lattice depth quench (\textit{L}-quench) (at $q=85$~Hz) and a quadratic Zeeman shift quench (\textit{Q}-quench) (to $q=44$~Hz) respectively.}
    \label{fig:cost_functions}
\end{figure}
This section discusses the optimization procedure used to determine $U_2$ and $V_2$ from simulation data with the two body and three body Hamiltonians in Eq. \ref{TwoBody} and \ref{ThreeBody} of the main text. The accuracy of the results is assessed via the distance (cost) $D = \frac{1}{N_p} \sum_{f=f_{\rm low}}^{f_{\rm high}} |A^{\rm exp}(f) - A^{\rm theory}(f)|$ between the theory and experimental spectral curves, where $N_p$ is the number of points in the summation. To isolate the effect of two-body interactions, $U_2$ is computed purely by performing this minimization using $n=2$ data over a frequency range $f \in [f_{\rm low}, f_{\rm high}]$ in the vicinity of  $f_2^{\rm harm}$ based on the harmonic theory prediction for $U_2$. In practice, we take $f_{\rm low / high } = f^{\rm harm}_2 \pm 2 \Delta f$, where $\Delta f =  (t_{\rm final} - t_{\rm init})/2\pi$ is the frequency  resolution. Note that we use zero padding to generate `smooth' Fourier spectra $A(f)$, discretized on a much finer grid than $\Delta f$. Fig.~\ref{fig:cost_functions}(a) and (b) show $D$ vs $U_2$ for both types of quench sequence. We find that $D$ is typically a smooth function of $U_2$, with a clear minimum at an optimal $U_2$ value. The curve is well approximated by a Gaussian distribution fit (dotted line) which can be used to estimate the uncertainty on $U_2$.
In order to reduce the impact of the chosen time interval on the Fourier transform, the optimization procedure is carried out at least 10 times, each with a slightly different time interval. The results shown in Fig.~\ref{fig:cost_functions} correspond to the $D$ value obtained after averaging over the $D$ values obtained from the different simulation results. The uncertainties stated in the main text are similarly derived.   
     
The optimization of $V_2$ is more complicated. Due to the finite time interval of the experimental signal, the sampling resolution in the frequency space ($\approx 15$ Hz) is larger than the expected shift in $f_3$ due to three-body interactions. However, higher atom numbers can have substantially larger shifts, scaling with $n^3$. We therefore perform independent simulations for each $n$ value and consider the linear superposition of these signals, i.e. $A^{\rm theory}(f) = \sum^{n_{\rm cut}}_{n=2} \chi_n A^{\rm theory}_n(f),$ up to a maximum atom number $n_{\rm cut} = 8$ selected to bound the dominate Mott lobe occupations observed in prior  works~\cite{Zihe2019,Jared2021Manipulation}. 
The relative population in the $n=1$ Mott lobe cannot be inferred using this method as it does not contribute to the spin dynamics. Therefore, for the results shown in Fig.~\ref{fig:4}  of the main text we have normalized $\chi_n$ such that $\sum_{n=2}^{n_{\rm cut}} \chi_n  = 1$.
We select $f_{\rm low}$ and $f_{\rm high}$ to encompass all of the band-gap frequencies shown in Fig.~\ref{fig:1} of the main text. However, frequencies above $f_{\rm high}$ are also included, albeit with a weighted contribution to  $D$ via a scale factor of 0.3, up to approximately 1000~Hz. This helps to avoid the over-estimation of the contribution from high Mott lobes and the excited states of lower Mott lobes ($n \approx n_{\rm cut}$), i.e. a large $\chi_3$ may assist in matching an experimental frequency, but this may also introduce weight in high frequencies that are undesirable. While typically the eigenmodes associated with these higher frequencies have less overlap with the initial state, 
this is not always true. 

We minimize $D$ to determine the optimal values of $V_2$ and $\chi_n$ simultaneously. The optimization of $\chi_n$ is carried out with the Nelder-Mead algorithm using the Optim package~\cite{Mogensen2018} in Julia for a given $V_2$. This allows us to obtain $D$ as a function of $V_2$ with optimal $\chi_n$ choices. Fig.~\ref{fig:cost_functions} (c) and (d) show $D$ vs $V_2$ for the same datasets used for Fig.~\ref{fig:cost_functions}(a) and (b) respectively. While the cost landscape is more complex than the case for $U_2$, it still has a clear minimum. Once again, fitting a Gaussian distribution to the curve near the minima allows us to estimate the uncertainty. In cases where the curve near the minima is relatively flat and not entirely convex (though the curve outside this region is typically smooth and convex), we allow the center of the Gaussian to also be a fitting parameter. The off-set of this center from the true minimum is then added to the standard deviation of the Gaussian to give the final uncertainty estimate.  

\begin{figure}[b]
    \centering
 \includegraphics[width=\linewidth]{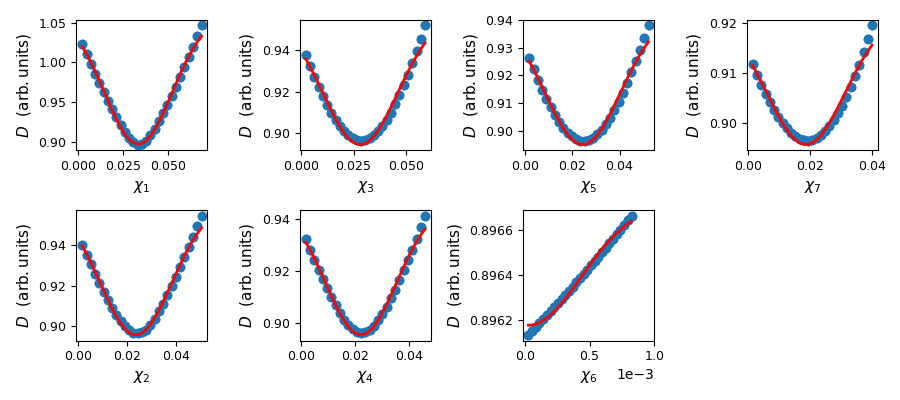}
    \caption{Cost function vs $\chi_n$ for all $n\in \{2, 8$\}  in the case of a quench at $q/U_2 \approx 0.85$. Points (blue) indicate data while the curves (red) are a Gaussian fit to the data. For the lower-right plot $\chi_6^{\rm opt}\approx 0$ and the Gaussian curve is not a good description.} 
    \label{fig:error_chi}
\end{figure}

While the aforementioned procedure determines the optimal $\chi_n$, the uncertainty of these parameters in principle requires determining $D$ in an $n_{\rm cut}-1$ dimensional space; information that is not directly obtained by the Nelder-Mead algorithm. As such, we obtain estimates for the uncertainty by looking at how $D$ varies as a function of a single $\chi_{n'}$, with all other  $\chi_n$ ($n \neq n'$) fixed to their optimal values $\chi^{\rm opt}_n$, i.e. a one dimensional cut of the full landscape. Fig.~\ref{fig:error_chi} shows $D$ vs $\chi_n$ for all $n$ in the case of a quench at $q/U_2\approx 0.85$. It can be seen that the cost functions are smooth and well described by a Gaussian fit with standard deviation $\sigma_n$ (red line), which provides an estimate for the uncertainty. For simplicity, this procedure is carried out with un-normalized values of $\chi_n$, as this allows the distribution to be symmetric and therefore better described by the fit. In cases where $\chi_n \approx 0 $ the Gaussian fit is not a good approximation, and we refrain from estimating an error. This tends to occur only when a given $A_n(f)$ is incapable of approximating the data. An example is given in the lower-right plot for $\chi_6$. Once $\sigma_{n'}$ is determined we calculate a lower and upper bound $\chi^{\rm \pm}_{n'}= \chi^{\rm opt}_{n'} \pm \sigma_{n'}$. Finally, we re-calculate the normalization  at each of these bounds, $\mathcal{N}^{\rm \pm}_{n'} = \sum^{n_{\rm cut}}_{n\neq n'} \chi^{\rm opt}_n + \chi_{n'}^{\rm \pm},$ which defines the uncertainty for the normalized parameters as $\Delta \chi^{\pm}_{n'}= \sigma_{n'}/\mathcal{N}^{\rm \pm}_{n'}.$ This is the uncertainty shown with the error bars in Fig.~\ref{fig:4} of the main text. Note that this approximates the naive estimate of $\sigma_{n'} / \sum_{n=2}^{\rm n_{cut}} \chi_n$ when the uncertainty is small ($\sigma_n \ll \chi_n$).


\subsection{Example spectral densities $A_n(f)$}
\begin{figure}[p]
	\includegraphics[width=13.2cm]{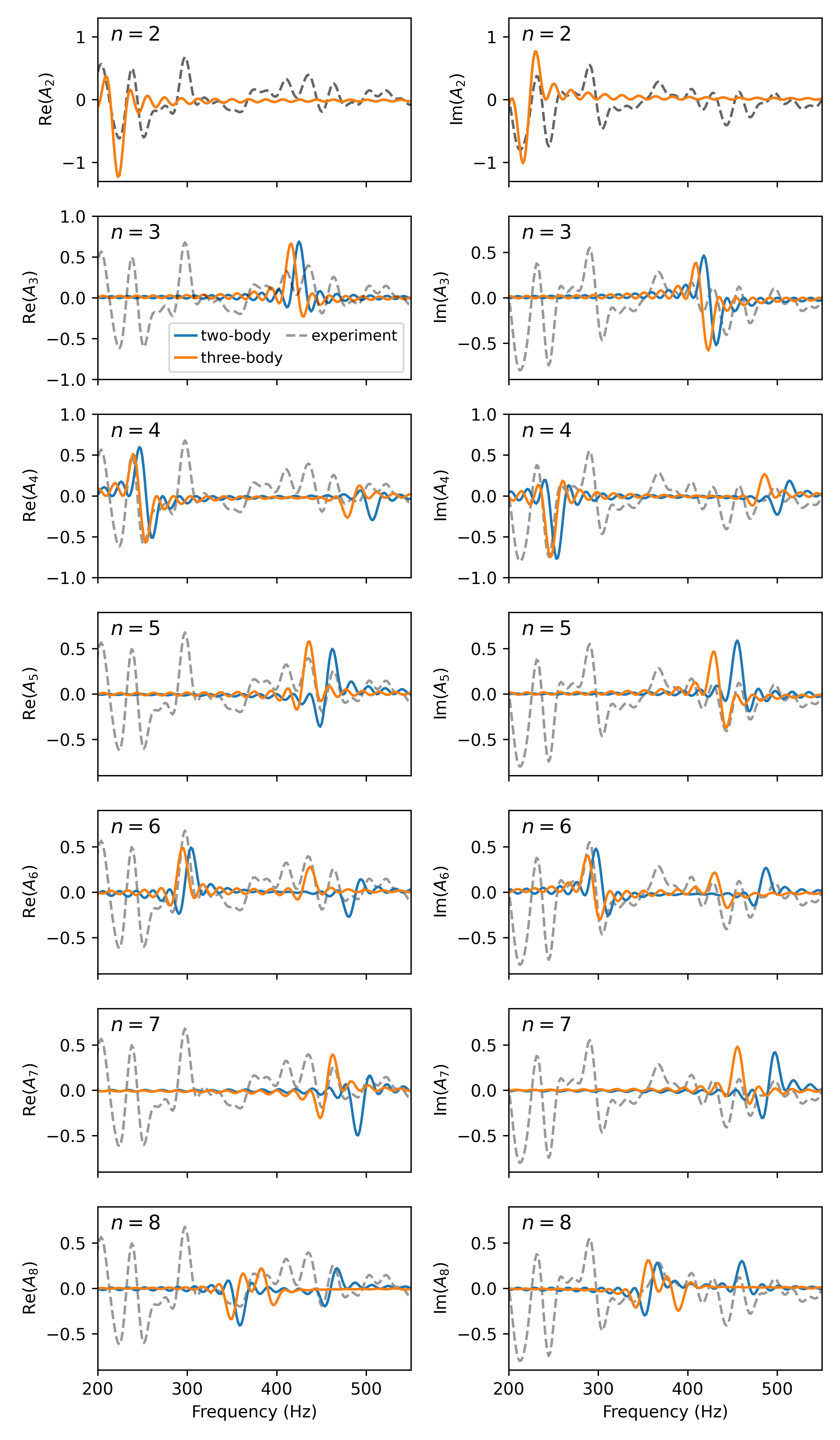}
	\caption{Real (left column) and imaginary (right column) parts of the spectral density $A_n$ for the quench with $q/U_2 \approx 0.85$. Every row corresponds to the simulation results with a different number of particles $n$. The full experimental result (faded dotted line) is included to guide the eye to important frequencies. We arbitrarily normalize the results for ease of visual comparison with the experimental results, though the two-body and three-body results are treated similarly. The results are only displayed over the frequency range of interest.}
	\label{fig:65_allN}
\end{figure}

\begin{figure}[p]
	\includegraphics[width=13.2cm]{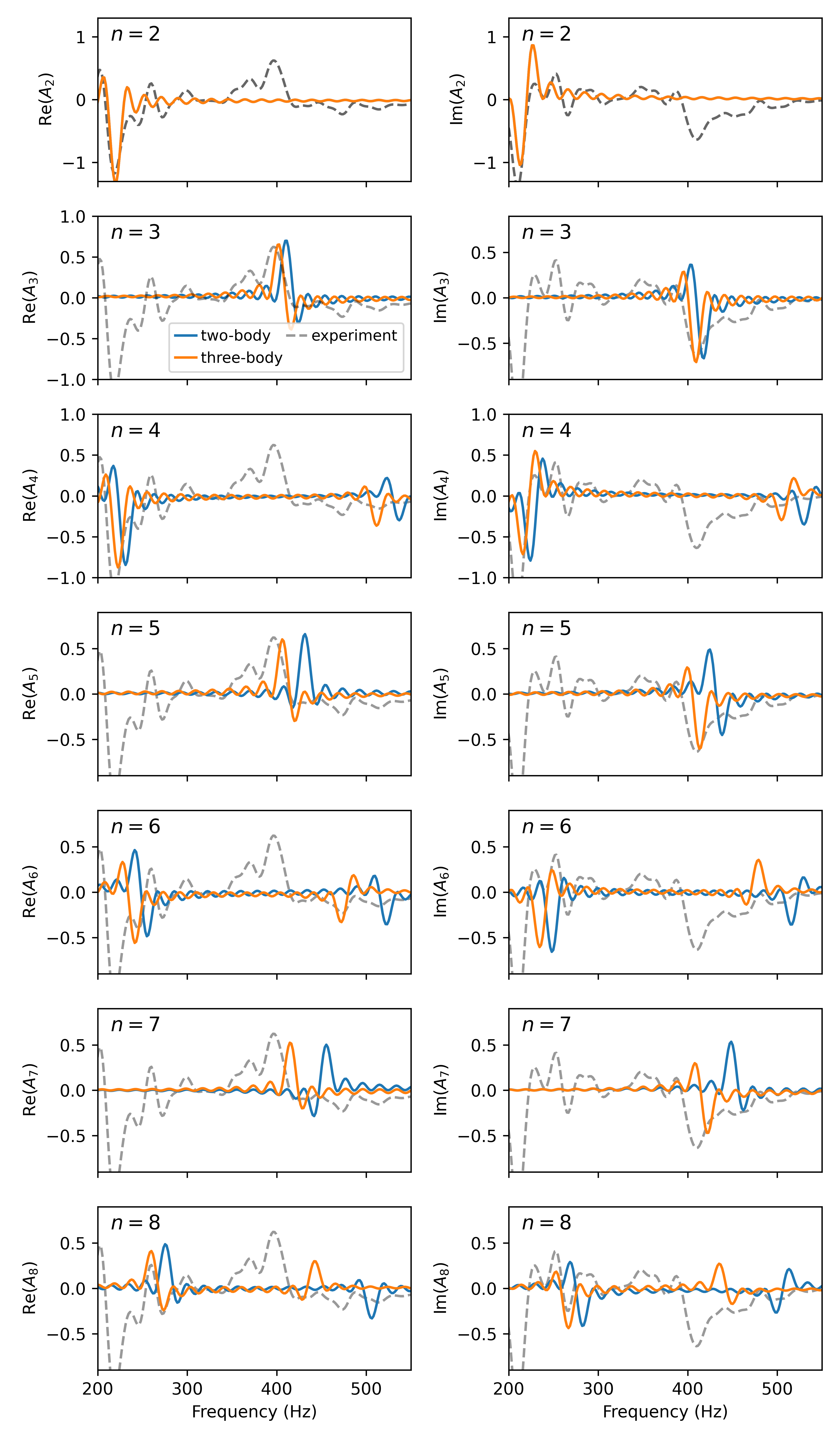}
	\caption{Real (left column) and imaginary (right column) parts of the spectral density $A_n$ for the quench with $q/U_2 \approx 0.60$. Every row corresponds to the simulation results with a different number of particles $n$. The full experimental result (faded dotted line) is included to guide the eye to important frequencies. We arbitrarily normalize the results for ease of visual comparison with the experimental results,  though the two-body and three-body results are treated similarly. The results are only displayed over the frequency range of interest.}
	\label{fig:44_allN}
\end{figure}

Examples of the individual signal contributions from each $n$ simulation are shown in Fig.~\ref{fig:65_allN} ($q/U_2 \approx 0.85$) and in Fig.~\ref{fig:44_allN}  ($q/U_2 \approx 0.60$) for the quenches shown in Fig.~\ref{fig:4} of the main text. Data for both the real (left column) and imaginary (right column) parts of the spectral density are displayed. For a given $n$, it can be seen that the three-body and two-body interactions have peaks which are slightly offset, in conformity with Fig.~\ref{fig:1}(b) of the main text. While we restrict the plots to the most relevant frequency range, in some instances higher frequency oscillations are also visible.  The aforementioned fitting procedure does not neglect these contributions. For some frequencies, the experimental signal can only be explained by a single $n$ value. However, where different $n$ have comparable frequencies, interference may occur between the different signals in the superposition. Signals that do not contribute towards a viable approximation of the experimental result will be weighted by a very small $\chi_n$ value.   

\subsection{Oscillation frequencies from quantum theory} \label{sec:oscillation_quantum}
The quantum state at time $t$ is expressed in terms of the eigenmodes $\ket{\nu}$ and eigenvalues $\epsilon_v$ as 
\begin{align}
    \ket{\psi(t)} = \sum_{v} c_v e^{-i \epsilon_v t} \ket{v}.
\end{align}
where $c_v = \braket{\nu}{\psi(0)}$. An arbitrary observable $\mathcal{O}(t) = \bra{\psi(t)}\hat{O}\ket{\psi(t)}$ can be easily expressed in terms of the eigenmodes as 
\begin{align}
 \mathcal{O}(t) 
 =  \sum_{\nu } |c_{\nu}|^2 \bra{\nu}  \hat{O} \ket{\nu}   +  \sum_{\mu > \nu} 2 \text{Re} \Big( c_{\nu} c^*_{\mu}  \bra{\mu}  \hat{O} \ket{\nu}    e^{-i (\epsilon_{\nu} - \epsilon_{\mu} )t} \Big) ,
\end{align}
where all the time-dependence is contained in the second term on the right, which contains the off-diagonal matrix elements. Therefore, the deviation $\delta \mathcal{O}(t) =\mathcal{O}(t) - \bar{\mathcal{O}} $ from the time-averaged value $\bar{\mathcal{O}}$ is given by
\begin{align}
    \delta \mathcal{O}(t) = \sum_{\mu > \nu}   \Big( 2 \text{Re} \big( A_{\mu \nu} \big)  \cos \{(\epsilon_{\nu} - \epsilon_{\mu} )t\} + 2 \text{Im}\big(A_{\mu \nu}\big)   \sin \{(\epsilon_{\nu} - \epsilon_{\mu} )t\}  \Big) 
\end{align}
where $A_{\mu \nu} = c_{\nu} c^*_{\mu}  \bra{\mu} \hat{O} \ket{\nu}$. Taking the Fourier transform $  \delta \mathcal{O}(\omega) = \frac{1}{2\pi}\int^{\infty}_{-\infty} dt e^{i \omega t}  \delta \mathcal{O}(t)$ gives
\begin{align}
    \delta \mathcal{O}(\omega) = \sum_{\mu > \nu}   \Big\{ \text{Re} \big( A_{\mu \nu} \big)   \Big( \delta(\omega - \epsilon_{\nu} + \epsilon_{\mu} ) + \delta(\omega + \epsilon_{\nu} - \epsilon_{\mu} ) \Big)  +  i  \text{Im}\big(A_{\mu \nu}\big) \Big( \delta(\omega - \epsilon_{\nu} + \epsilon_{\mu} )   + \delta(\omega + \epsilon_{\nu} - \epsilon_{\mu} )  \Big)  \Big\} 
\end{align}
Considering only the solution for positive frequencies, the real and imaginary parts of the Fourier transform are
\begin{align}
    {\rm Re}[ \delta \mathcal{O}(\omega)] & =  \sum_{\mu > \nu}    \text{Re} \big( A_{\mu \nu} \big)  \delta(\omega - |\epsilon_{\nu} + \epsilon_{\mu}|) , \\
    {\rm Im}[ \delta \mathcal{O}(\omega)] & =  \sum_{\mu > \nu}    \text{Im} \big( A_{\mu \nu} \big)  \delta(\omega - |\epsilon_{\nu} + \epsilon_{\mu}|) 
\end{align}
This illustrates that the spectra in Fig.~\ref{fig:2} of the main text, which show the real and imaginary parts of the Fourier transform of $\rho_0$, contain information about the off-diagonal matrix elements and initial state overlap factors. 

%